\documentclass[10pt,sort&compress]{article}
\usepackage{graphicx,rotating,multirow}
\usepackage{bm,multirow}
\usepackage{amsmath,amssymb,color,mathrsfs}

\usepackage{footnote}
\usepackage{citesort}
\usepackage{cuted}

\def\lambdabar{\lambda\kern-1ex\raise0.55ex\hbox{--}}

\newcommand{\AJM}{{\em Am. J. Math.} }
\newcommand{\AJP}{{\em Am. J. Phys.} }
\newcommand{\AdM}{{\em Adv. Math.} }
\newcommand{\AM}{{\em Annals Math.} }

\newcommand{\APB}{{\em Ann. Phys. (Berlin)} }
\newcommand{\APNY}{{\em Ann. Phys. (N.Y.)} }

\newcommand{\CMP}{{\em Commun. Math. Phys.} }

\newcommand{\CPL}{{\em Chem. Phys. Lett.} }
\newcommand{\CRA}{{\em C. R. Acad. Sci. Ser. A} }
\newcommand{\EJP}{{\em Eur. J. Phys.} }

\newcommand{\EPL}{{\em Europhys. Lett.} }

\newcommand{\IJQC}{{\em Int. J. Quantum Chem.} }

\newcommand{\JCP}{{\em J. Chem. Phys.} }

\newcommand{\JHEP}{{\em J. High Energy Phys.} }
\newcommand{\JMC}{{\em J. Math. Chem.} }
\newcommand{\JMP}{{\em J. Math. Phys.} }
\newcommand{\jpa}{{\em J. Phys. A} }

\newcommand{\JSP}{{\em J. Stat. Phys.} }
\newcommand{\LMP}{{\em Lett. Math. Phys.} }

\newcommand{\NJP}{{\em New J. Phys.} }

\newcommand{\NP}{{\em Nature Phys.} }
\newcommand{\PA}{{\em Physica A} }

\newcommand{\PJP}{{\em Pramana-J. Phys.} }
\newcommand{\PLA}{{\em Phys. Lett. A} }

\newcommand{\PRA}{{\em Phys. Rev. A} }
\newcommand{\PRB}{{\em Phys. Rev. B} }
\newcommand{\PRD}{{\em Phys. Rev. D} }
\newcommand{\PRE}{{\em Phys. Rev. E} }
\newcommand{\PRL}{{\em Phys. Rev. Lett.} }
\newcommand{\PRe}{{\em Phys. Rep.} }

\newcommand{\RMP}{{\em Rev. Mod. Phys.} }

\newcommand{\SR}{{\em Sci. Rep.} }

\definecolor{officegreen}{rgb}{0,0.5,0}
\definecolor{pakistangreen}{rgb}{0,0.4,0}
\definecolor{palatinatepurple}{rgb}{0.41,0.16,0.38}
\definecolor{sangria}{rgb}{0.57,0,0.04}

\begin{document}
\title{Quantum information measures of the one-dimensional Robin quantum well}
\author{O. Olendski\footnote{Department of Applied Physics and Astronomy, University of Sharjah, P.O. Box 27272, Sharjah, United Arab Emirates; E-mail: oolendski@sharjah.ac.ae}}

\maketitle

\begin{abstract}
Shannon quantum information entropies $S_{x,k}$, Fisher informations $I_{x,k}$, Onicescu energies $O_{x,k}$ and statistical complexities $e^{S_{x,k}}O_{x,k}$ are calculated both in the position (subscript $x$) and momentum ($k$) representations for the Robin quantum well characterized by the extrapolation lengths $\Lambda_-$ and $\Lambda_+$ at the two confining surfaces. The analysis concentrates on finding and explaining the most characteristic features of these quantum information measures in the whole range of variation of the Robin distance $\Lambda$ for the symmetric, $\Lambda_-=\Lambda_+=\Lambda$, and antisymmetric, $\Lambda_-=-\Lambda_+=\Lambda$, geometries. Analytic results obtained in the limiting cases of the extremely large and very small magnitudes of the extrapolation parameter are corroborated by the exact numerical computations that are extended to the arbitrary length $\Lambda$. It is confirmed, in particular, that the entropic uncertainty relation $S_{x_n}+S_{k_n}\geq1+\ln\pi$ and general inequality $e^SO\geq1$, which is valid both in the position and momentum spaces, hold true at any Robin distance and for every quantum state $n$. For either configuration, there is a range of the extrapolation lengths where the rule $S_{x_{n+1}}(\Lambda)+S_{k_{n+1}}(\Lambda)\geq S_{x_n}(\Lambda)+S_{k_n}(\Lambda)$ that is correct for the Neumann ($\Lambda=\infty$) or Dirichlet ($\Lambda=0$) boundary conditions, is violated. Other analytic and numerical results for all measures are discussed too and their physical meaning is highlighted.
\end{abstract}

\section{Introduction}\label{sec1}
Model of the infinitely deep quantum well (QW) serves as a test bed for checking both fundamental physical laws as well as their possible technological applications \cite{Belloni1}. For example, recent theoretical analysis \cite{AlHashimi1} proved that the Heisenberg uncertainty relation
\begin{equation}\label{Heisenberg1}
\Delta x\Delta k\ge\frac{1}{2},
\end{equation}
in general, does not hold for the one-dimensional (1D) structure with Robin boundary condition (BC) \cite{Gustafson1}
\begin{equation}\label{Robin1}
\left.{\bf n}{\bm\nabla}\Psi\right|_{\cal S}=\left.\frac{1}{\Lambda}\Psi\right|_{\cal S}
\end{equation}
with $\bf n$ being an inward unit normal to the surface ${\cal S}$. In equations above, $\Delta x$ and $\Delta k$ are, respectively, position and wave vector standard deviations:
\begin{subequations}\label{DeltaXK1}
\begin{eqnarray}\label{DeltaX1}
\Delta x&=&\sqrt{\left<x^2\right>-\left<x\right>^2}\\
\label{DeltaK1}
\Delta k&=&\sqrt{\left<k^2\right>-\left<k\right>^2},
\end{eqnarray}
\end{subequations}
where the associated moments $\left<x^n\right>$ and $\left<k^n\right>$, $n=1,2,\ldots$, are expressed through the corresponding position $\Psi(x)$ and momentum $\Phi(k)$ wave functions:
\begin{subequations}\label{XKaveraging1}
\begin{eqnarray}\label{Xaveraging1}
\left<x^n\right>&=&\int_{-d/2}^{d/2}x^n\rho(x)dx\\
\label{Kaveraging1}
\left<k^n\right>&=&\int_{-\infty}^\infty k^n\gamma(k)dk
\end{eqnarray}
\end{subequations}
with the densities
\begin{subequations}\label{densityXK_1}
\begin{eqnarray}\label{densityX_1}
\rho(x)&=&|\Psi(x)|^2\\
\label{densityK_1}
\gamma(k)&=&|\Phi(k)|^2,
\end{eqnarray}
\end{subequations}
and $d$ is the width of the well whose edges confine the motion inside the interval $-d/2\leq x\leq x/2$. Waveform $\Psi(x)$ satisfies the 1D analog of the general 3D Schr\"{o}dinger equation for the particle of the mass $m$ in the potential $V({\bf r})$
\begin{equation}\label{Schrodinger1}
-\frac{\hbar^2}{2m}{\bm\nabla}^2\Psi({\bf r})+V({\bf r})\Psi({\bf r})=E\Psi({\bf r}),
\end{equation}
and momentum counterpart $\Phi(k)$ is its Fourier transform:
\begin{equation}\label{FunctionPhi1}
\Phi(k)=\frac{1}{\left(2\pi\right)^{1/2}}\int_{-d/2}^{d/2}e^{-ikx}\Psi(x)dx.
\end{equation}
For our geometry, Eq.~\eqref{Schrodinger1} supplemented by the BC from Eq.~\eqref{Robin1} has a countably infinite number of solutions with the real position functions $\Psi_n(\Lambda;x)$ and associated energies $E_n(\Lambda)$ with the former ones being assumed orthogonalized as
\begin{subequations}\label{Normalization1}
\begin{align}\label{Normalization1_X}
\int_{-d/2}^{d/2}\Psi_n(\Lambda;x)\Psi_{n'}(\Lambda;x)dx&=\delta_{nn'},\quad n,n'=0,1,2,\ldots
\intertext{$\delta_{nn'}$ is Kronecker delta. Accordingly, momentum functions are orthonormalized too:}
\label{Normalization1_K}
\int_{-\infty}^\infty\Phi_{n'}^\ast(\Lambda;k)\Phi_n(\Lambda;k)dk&=\delta_{nn'}.
\end{align}
\end{subequations}
Real value of the coefficient $\Lambda$ that regulates linear relation between the position function $\Psi(x)$ and its spatial derivative $\Psi'(x)$  at the interfaces warrants that no current with the density $\bf j$ flows through the surfaces  $x=\pm d/2$:
\begin{equation}\label{CurrentDensity2}
\left.{\bf nj}\right|_{\cal S}\equiv0\quad{\rm at}\quad{\rm Im}(\Lambda)=0,
\end{equation}
what can be easily shown from the corresponding expression:
\begin{equation}\label{CurrentDensity1}
{\bf j}=-\frac{e\hbar}{m}{\rm Im}(\Psi^*{\bm\nabla}\Psi),
\end{equation}
$e$ is an absolute value of the electronic charge. It was proved \cite{AlHashimi1,Olendski2} that the Heisenberg relation, Eq.~\eqref{Heisenberg1}, holds for the Dirichlet BC, $\Psi|_{\cal S}=0$, only since it was assumed during its derivation that the wave function $\Psi$ vanishes at infinity what is not the case for the finite volume with the nonzero Robin length $\Lambda$. To account for the influence of the non Dirichlet, $\Lambda\neq0$, surfaces, one introduces the BC dependent terms into the Heisenberg relation; for example, for the lowest level of the Neumann QW, $\Psi'(\pm d/2)=0$, with its zero energy, $E_0(\infty)=0$, and constant position function, $\Psi_0(\infty;x)=d^{-1/2}$, the uncertainty correlation turns to $2mE_0\geq0$, which is indeed satisfied as an equality \cite{AlHashimi1}. Note that the higher lying  Neumann orbitals do obey the standard relation from Eq.~\eqref{Heisenberg1}. A violation of the Heisenberg uncertainty relation was experimentally demonstrated for a number of systems, e.g., in a neutron-optical settlement that records the error of a spin-component measurement as well as the disturbance caused on another spin-component \cite{Erhart1}. A lot of theoretical efforts was devoted to the improvement of the standard Heisenberg inequality, see, e.g., Refs.~\cite{Erhart1,Maccone1,Li1,Song1,Li2,Herdegen1} and literature therein. A different approach employs the properties of the other quantum information measures. For example, a sum of the position $S_x$ and momentum $S_k$ quantum information entropies that for our geometry are defined as
\begin{subequations}\label{Entropy1}
\begin{eqnarray}\label{Entropy1_X}
S_x&=&-\int_{-d/2}^{d/2}\rho(x)\ln\rho(x)dx\\
\label{Entropy1_K}
S_k&=&-\int_{-\infty}^{\infty}\gamma(k)\ln\gamma(k)dk
\end{eqnarray}
\end{subequations}
calculated for the same lowest Neumann level \cite{Olendski2,Bialynicki2} does satisfy the fundamental inequality
\begin{equation}\label{EntropicInequality1}
S_t\equiv S_x+S_k\ge1+\ln\pi
\end{equation}
that was rigorously proved for the arbitrary 1D geometry by I. Bia{\l}ynicki-Birula and J. Mycielski \cite{Bialynicki3} and W. Beckner \cite{Beckner1} with earlier conjectures from H. Everett \cite{Everett1} and I. I. Hirschman \cite{Hirschman1}. This example reconfirms \cite{Deutsch1,Partovi1} that entropic uncertainty relation, Eq.~\eqref{EntropicInequality1}, is stronger than its Heisenberg counterpart, Eq.~\eqref{Heisenberg1}, as it presents more general base for defining 'uncertainty' \cite{Bialynicki2,Wehner1,Coles1}. Accordingly, an investigation of the quantum entropies $S_x$ and $S_k$ is an important problem for the structures with the non Dirichlet BCs.

In the present research, an exact thorough examination of the different quantum information measures of the Robin QW is provided with the emphasis on their dependencies on the left $\Lambda_-$ and right $\Lambda_+$ extrapolation parameters, which, in general, might be different. Classical information entropy was introduced by C. E. Shannon for the mathematical analysis of communication as a measure "of information, choice and uncertainty" \cite{Shannon1}. Its quantum counterparts from Eqs.~\eqref{Entropy1} describe quantitatively the lack of our knowledge about position, Eq.~\eqref{Entropy1_X}, and momentum, Eq.~\eqref{Entropy1_K}, localization of the nano object: the greater the entropy is, the less we know about the corresponding property. This missing information is an inherent essential characteristics of any quantum motion; in particular, this uncertainty can not be eliminated by the increase of the accuracy of the measuring device. As Eq.~\eqref{EntropicInequality1} manifests, the Shannon entropies $S_x$ and $S_k$ of the quantum particle are not independent from each other: the more information we get about position of the corpuscle, the less we know about its momentum. From mathematical point of view, due to the presence of the logarithm, the functionals from Eqs.~\eqref{Entropy1} can take negative values; for example, for the position Shannon entropy it happens when the parts of the corresponding density, which are larger than unity, in their contribution to the integral from Eq.~\eqref{Entropy1_X} overweigh those with $\rho(x)<1$. Physically, the logarithm in the integrands leads to the dimensional problem; namely, the functionals from  Eqs.~\eqref{Entropy1} for the {\em continuous} distributions were obtained from their {\em discrete} counterpart; viz., for the complete set of all $N$ possible distinct events with their probabilities $0\leq p_n\leq1$, $n=1,2,\ldots, N$, such that $\sum_{n=1}^Np_n=1$, one defines the entropy as \cite{Shannon1}:
\begin{equation}\label{ShannonDiscrete1}
S=-\sum_{n=1}^Np_n\ln p_n,
\end{equation}
which is, obviously, dimensionless. In turn, the Shannon entropies for the continuous distributions are measured, as it follows from Eqs.~\eqref{Entropy1}, in units of the logarithm of the length what is physically ambiguous. The first remedy to cure this longstanding equivocation of the dimension of the entropies is to exponentiate  \cite{Srinivas1} the expressions from Eqs.~\eqref{Entropy1} arriving in this way at the Shannon information-theoretic position and momentum lengths. The second widely used method for finding the purport of the expressions $\ln\rho$ and $\ln\gamma$ is to measure distances in units of some characteristic length of the system under consideration \cite{Dodonov1} making in this way the corresponding densities dimensionless whereas the entropies in this case do depend on the choice of this unit; for example, below all lengths are expressed in units of the well width $d$. The very recent approach proposes to use as an argument of the logarithm the position density normalized to its maximal value what makes this modified dimensionless entropy strictly positive \cite{Flores1}. It is important to underline that despite the ambiguity of the dimensions of the position and momentum entropies, their sum is a scaling-independent dimensionless quantity, as it is seen, for example, from Eq.~\eqref{EntropicInequality1}. Apart from this fundamental relation, the rapidly growing interest in the study of the entropies $S_x$ and $S_k$ for different structures of miscellaneous dimensionality is stimulated by the fact that with the help of this analysis many other important physical and chemical quantities and phenomena can be understood; for example, to name just a few, entropy maximization was used for constructing of Compton profiles for helium and atomic and molecular hydrogen \cite{Sears2}; within the local plasma approximation, the mean excitation energies of a series of atoms and molecules were computed from the position Shannon entropies \cite{Ho1}; position and momentum functionals were employed in phenomenological description of the transition state, the bond breaking and bond forming processes of some elementary chemical reactions \cite{Esquivel2}; etc.

Below, position
\begin{subequations}\label{Fisher1}
\begin{align}
\label{Fisher1_X}
I_x&=\int_{-d/2}^{d/2}\rho(x)\left[\frac{d}{dx}\ln\rho(x)\right]^2dx=\int_{-d/2}^{d/2}\frac{\rho'(x)^2}{\rho(x)}\,dx\\
\intertext{and momentum}
\label{Fisher1_K}
I_k&=\int_{-\infty}^\infty\gamma(k)\left[\frac{d}{dk}\ln\gamma(k)\right]^2dk=\int_{-\infty}^\infty\frac{\gamma'(k)^2}{\gamma(k)}dk
\end{align}
\end{subequations}
Fisher informations \cite{Fisher1} of the Robin QW are calculated too. The functional of the form from Eqs.~\eqref{Fisher1} was originally introduced in 1925 by the English statistician and biologist whose name it now bears as "a measure of the intrinsic accuracy of an error curve" and was "conceived as the amount of information in a single observation" \cite{Fisher1} belonging to a corresponding distribution. As expressions from Eqs.~\eqref{Fisher1} manifest, Fisher information can not take negative values as it is a mean value of the square of the relative speed of change of the corresponding probability. Derivatives of the densities $\rho(x)$ and $\gamma(k)$ in Eqs.~\eqref{Fisher1} make the Fisher information a local measure of uncertainty whereas the quantum entropies with the logarithms in their integrands are global descriptions of the charge delocalization. In the density functional theory (DFT), the position Fisher information defines the kinetic energy of the many-particle system what allows to reformulate the quantum mechanical variation principle as a principle of minimal information \cite{Sears1} thus establishing the link between DFT and information theory. Use of the Fisher information as an efficient estimator of the physical parameters encoded in a set of quantum states and as a measure of system disorder far transcends its applications in physics and is expanded in many other branches of science such as, for example, genetic evolution, macroeconomics and cancer growth \cite{Frieden1}.

In addition to the quantum Shannon entropies and Fisher informations, position $O_x$ and momentum $O_k$ Onicescu energies \cite{Onicescu1}
\begin{subequations}\label{Onicescu1}
\begin{eqnarray}\label{Onicescu1_X}
O_x&=&\int_{-d/2}^{d/2}\rho^2(x)dx\\
\label{Onicescu1_K}
O_k&=&\int_{-\infty}^\infty\gamma^2(k)dk
\end{eqnarray}
\end{subequations}
are analyzed as functions of the Robin distance too. Eqs.~\eqref{Onicescu1} show that the strictly positive quantities $O_x$ and $O_k$ are measured in units of the inverse volume of the field upon which they are calculated and represent the mean values of the corresponding probability densities or quadratic deviations from the probability equilibria (disequilibria); in other words, they carry information on how close the dependencies $\rho(x)$ and $\gamma(k)$ are to their uniform counterparts. Already forty  years ago it was noted \cite{Hyman1} that for the atom a position disequilibrium, which can be called 'average electron density', is an experimentally measurable quantity related to the X-ray intensity scattered by the element. It also has interesting relationships to other physically important quantities \cite{Tao1}. Contrary to the Shannon entropy with its fundamental inequality, Eq.~\eqref{EntropicInequality1}, similar universal restrictions for the Fisher information or Onicescu energy are not known though some lower (for $I_xI_k$ \cite{Stam1,Dembo1,Romera1,Dehesa2,Dehesa3}) or upper (for $O_xO_k$ \cite{Ghosal1}) bounds involving the products of position and momentum components have been obtained for several particular systems. An interesting history of the discussion on the universality of the 1D Fisher uncertainty relation
\begin{equation}\label{FisherUncertainty1}
I_xI_k\geq4
\end{equation}
can be found in Refs.~\cite{Plastino1,Saha1}.

Three types of functionals introduced in Eqs.~\eqref{Entropy1}, \eqref{Fisher1} and \eqref{Onicescu1}, describe different features of the particle distribution, namely, its spreading, its oscillation structure and departure from equilibrium, respectively \cite{Esquivel1}. Since these properties are independent of each other, their appropriate combinations can reflect two different aspects of the probability density \cite{Esquivel1}; for example, for the discrete field with $N$ events the entropy (information energy) reaches maximum of $\ln N$ (minimum of $1/N$) when the likelihoods of all occurences are equal, $p_n=1/N$, $n=1,2\ldots,N$, whereas the zero minimum (unit maximum) takes place with the probability of one event being certain with all others turning to zeros, $p_n=\left\{\begin{array}{cc}
1,&n=m\\
0,&n\neq m
\end{array}\right.$, $m=1,2\ldots,N$ \cite{Chatzisavvas1}. Thus, it is natural to study the statistical measure of complexity \cite{Catalan1}
\begin{equation}\label{CGLdefinition1}
CGL=e^SO,
\end{equation}
which, referring to our earlier discussion on units of measuring the Shannon entropies, is a dimensionless scaling-independent quantity. It helps to detect and analyze not only the randomness represented by the first multiplier in the right-hand side of Eq.~\eqref{CGLdefinition1}  but also the structure of the probability density described by the factor $O$. The above example can be easily extended to the continuous case; namely, for the uniform distribution the Onicescu energy is minimal and, since this configuration contains no any information about particle localization, the entropy is maximal. Any deviation from the equiprobability increases the disequilibrium and simultaneously provides some knowledge about corpuscle whereabouts thus decreasing the Shannon entropy. The positive or negative change of the complexity $CGL$ shows then which contribution to it prevails over the second measure. In the opposite limit of highly nonuniform distribution, the shape of the density approaches a Dirac $\delta$-function what results in the infinite value of $O$ describing a maximal departure from homogeneity whereas the entropy $S$ will tend to the unrestrictedly large negative values corresponding to a precise knowledge about the system. The value of the complexity is determined in this asymptotic case by the path along which the density turns into the $\delta$-function. From the very general principles, it can be shown that for any $l$-dimensional space (with arbitrary positive integer $l$) either position or momentum component of this product satisfies the inequality \cite{LopezRosa2}
\begin{equation}\label{CGLinequality1}
CGL\geq1,
\end{equation}
which is saturated only for the uniform distribution with a finite volume support what for our geometry corresponds to the position component of the lowest Neumann orbital. In addition, the complexity stays invariant under scaling, translation and replication \cite{Catalan1}. Analysis of the position and momentum components of this estimator applied to the neutral atoms with the nuclear charge $Z$ in the range $1\leq Z\leq103$ revealed a strong correlation between the shell-filling process in atomic systems and location of the $CGL$ extrema on the $Z$ axis \cite{LopezRosa2}. The complexity also appeared to be very useful in the analysis of the dynamics of brain electric activity \cite{Rosso1}.

Our discussion focuses on the analytic and numerical description of the position and momentum components of these three quantum-information measures together with the complexity $CGL$ for the two BC geometries: the first configuration is characterized by the same value of the extrapolation lengths on both confining interfaces, $\Lambda_-=\Lambda_+\equiv\Lambda$, and the other one  exhibits opposite signs of the same magnitudes of the Robin distance on the surfaces, $\Lambda_-=-\Lambda_+\equiv\Lambda$. It is shown, in particular, that for the symmetric configuration a sum of the two entropies $S_{t_n}(\Lambda)=S_{x_n}(\Lambda)+S_{k_n}(\Lambda)$ [a product of the two Onicescu energies $O_{x_n}(\Lambda)O_{k_n}(\Lambda)$] for each quantum level $n$ has as a function of the Robin distance $\Lambda$ a global minimum (maximum) at $\Lambda=0$. From physical meaning of the Shannon entropy and Onicescu energy one concludes then that  the Dirichlet BC provides the largest possible amount of information about simultaneous knowledge of the position and momentum while their overall disequilibrium is the biggest at this surface requirement too. For the two lowest split-off levels existing at the negative Robin length only with their energies unrestrictedly decreasing at the vanishing $|\Lambda|$, the associated position (momentum) Shannon entropies diverge in the same limit as negative (positive) $\ln|\Lambda|$ what results in a finite sum $S_{t_n}$ obeying, of course, at any extrapolation distance inequality~\eqref{EntropicInequality1}. Corresponding position (momentum) Onicescu energy for the asymptote $\Lambda\rightarrow-0$ diverges as $1/(2|\Lambda|)$ [turns to zero as $3|\Lambda|/(4\pi)$] keeping the product $O_{x_n}O_{k_n}$ bound at arbitrary length. Analytic expressions for the position and momentum Fisher informations are derived and analyzed in the whole range of the Robin distances. A remarkable feature of the antisymmetric geometry, $\Lambda_-=-\Lambda_+=\Lambda$, is the fact that energies $E_n$ and position quantum-information measures $S_{x_n}$, $I_{x_n}$ and $O_{x_n}$ of the excited levels, $n\geq1$, do not depend on the extrapolation length even though the corresponding functions $\Psi_n(\Lambda;x)$ do transform from the Dirichlet  to Neumann (at $\Lambda=\infty$) waveforms. For either configuration, there is a range of the Robin lengths where the rule $S_{t_{n+1}}(\Lambda)\geq S_{t_n}(\Lambda)$ that is correct for the Neumann or Dirichlet BCs, does not hold. It is also shown that inequality~\eqref{CGLinequality1} is always satisfied.

Structure of presentation below is as follows: a brief introduction into the problem is provided in Sec.~\ref{Sec_Formulation1} where, as a prerequisite to further discussion, energy spectra and position wave functions are considered too. Chapter~\ref{Sec_Symmetric1} is devoted to the analysis of the symmetric BCs with separate subsections presenting main analytic results, Subsec.~\ref{SubSec_AnalyticSymmetric}; momentum waveforms, Subsec.~\ref{SubSec_EnergySymmetric}, whereas Shannon entropy, Onicescu energy (together with the complexity $CGL$) and Fisher information being discussed in Subsecs.~\ref{SubSec_Shannon}, \ref{SubSec_Onicescu} and \ref{SubSec_Fisher}, respectively. Next, QW with the opposite positive and negative extrapolation lengths at the two confining interfaces is described in Sect.~\ref{Sec_Asymmetric}. Similar to the geometries studied before \cite{Olendski2,Olendski22,Olendski3,Olendski33}, a companion paper \cite{Olendski44} calculates statistical properties of the structures.

\section{Formulation, energy spectrum and position waveforms}\label{Sec_Formulation1}
Structures with Robin BC are ubiquitous in nature. A correct understanding of the processes taking place in acoustics \cite{Felix1}, electrodynamics \cite{Balian1,Katsenelenbaum1}, plasma \cite{Silva1}, scalar field theory with special attention to the Casimir effect \cite{Solodukhin1,Saharian1,Romeo1,Elizalde1,Teo1,Nazari1}, superconductivity \cite{Fink1,Montevecchi1,Kozhevnikov1,Giorgi1} where the coefficient $\Lambda$ is called the de Gennes distance \cite{deGennes1}, and other branches of physics and related fields \cite{Sapoval1,Essert1} inevitably requires finding solutions of the wave equation in the form of Eq.~\eqref{Schrodinger1} with the additional requirement from Eq.~\eqref{Robin1} due to the confining surface. A continuous variation of the extrapolation length from its zero magnitude to the extremely large positive values smoothly transforms the BC from the Dirichlet to the Neumann case what has its consequences, for example, in the change of the sign of the Casimir force between two parallel plates \cite{Teo1}. The most interesting is the situation when the coefficient $\Lambda$ takes negative values what physically means that the interface attracts the particle with its wave function being localized near the surface and the corresponding energy falling below zero. Structures with the negative de Gennes distance were experimentally realized with the help of superconductors \cite{Fink1,Kozhevnikov1} where it leads to the enhancement of the critical temperature. Reviews on the research on the structures obeying the BC from Eq.~\eqref{Robin1} can be found in Refs.~\cite{Olendski1,Olendski4,Olendski5,Grebenkov1}.

Considering quite long interest in the Robin structures, it is rather surprising that the results of their quantum information-theoretical analysis are very scarce. In fact, the previous contribution of the author \cite{Olendski3}, which describes all three measures of the single Robin wall in the transverse electric field $\mathcal{E}$, seems to be the only research on the subject so far. It is natural to expand this investigation to the two de Gennes walls, initially at $\mathcal{E}=0$. As a first step in pursuit of this endevour, below in the present section the eigen energies $E_n$ and functions $\Psi_n(x)$ of the position Shrodinger equation~\eqref{Schrodinger1} have been derived under the constraint imposed on $\Psi_n(x)$ in the form of Eqs.~\eqref{BC1}. Knowledge of these waveforms is used in the two subsequent sections for the direct calculation of the corresponding position measures $S_x$, $I_x$, $O_x$ and $CGL_x$. Next, for calculating their counterparts $S_k$, $I_k$, $O_k$ and $CGL_k$, momentum functions $\Phi_n(k)$ are obtained with the help of Eq.~\eqref{FunctionPhi1}. Throughout the whole research, special attention is paid to finding analytic expressions for all involved quantities and asymptotic cases of the large and small magnitudes of the extrapolation length when further formulae simplifications are possible.

1D Robin QW of the width $d$ assumes the free particle motion in the region $-d/2\leq x\leq d/2$ while at its edges the position waveform $\Psi(x)$ satisfies the BCs
\begin{subequations}\label{BC1}
\begin{eqnarray}\label{BC1_-}
\left[\frac{d\Psi(x)}{dx}-\frac{1}{\Lambda_-}\Psi(x)\right]_{x=-d/2}=0\\
\label{BC1_+}
\left[\frac{d\Psi(x)}{dx}+\frac{1}{\Lambda_+}\Psi(x)\right]_{x=+d/2}=0
\end{eqnarray}
\end{subequations}
with, in general, different Robin lengths $\Lambda_-$ and $\Lambda_+$. It is convenient from the very beginning to switch to the dimensionless units when all distances are measured in units of the well width $d$. Accordingly, the energies will be scaled in units of the ground state energy of the Dirichlet QW $\pi^2\hbar^2/(2md^2)$. Then, the transcendental equation for finding eigenenergies of the Hamiltonian reads \cite{Olendski1}:
\begin{equation}\label{EigenValue1}
\left(\frac{1}{\Lambda_-}+\frac{1}{\Lambda_+}\right)\pi E^{1/2}\cos\pi E^{1/2}+\left(\frac{1}{\Lambda_-\Lambda_+}-\pi^2E\right)\sin\pi E^{1/2}=0.
\end{equation}
It immediately shows that for the opposite signs of the equal magnitudes of the extrapolation lengths (as mentioned above, we will call such a geometry an antisymmetric one and denote the corresponding quantities by the subscript '$A$') the energies are \cite{Olendski1}:
\begin{equation}\label{AsymmetricSpectrum1}
E_{A_0}=-\frac{1}{\pi^2\Lambda^2},\quad E_{A_n}=n^2,\,n=1,2,\ldots,\quad\Lambda_-=-\Lambda_+\equiv\Lambda.
\end{equation}
So, interaction of the two interfaces yields the Dirichlet spectrum supplemented by the BC split-off state whose  negative energy is equal to its counterpart of the single attractive wall \cite{Olendski3}. The emergence of the level with $E<0$ is caused by the negative extrapolation length when for $\Lambda_+$ approaching zero from the left the waveform, as is shown below, becomes more and more localized at the corresponding surface and the energy in the same limit falls down as $-\Lambda^{-2}$ what is true for any $l$-dimensional domain \cite{Lacey1,Lou1,Levitin1,Daners1,Colorado1,Pankrashkin1,Exner1,Freitas1}. Normalized to unity,
\begin{equation}\label{Normalization2}
\int_{-1/2}^{1/2}\Psi^2(\Lambda;x)dx=1,
\end{equation}
position wave functions read:
\begin{subequations}\label{AsymmetricFunctions1}
\begin{eqnarray}\label{AsymmetricFunctions1_0}
\Psi_{A_0}(\Lambda;x)&=&\frac{1}{\left(\Lambda\sinh\frac{1}{\Lambda}\right)^{1/2}}\exp\!\left(\frac{x}{\Lambda}\right)\\
\Psi_{A_{2n-1}}(\Lambda;x)&=&\sqrt{\frac{2}{1+\frac{1}{[\pi(2n-1)\Lambda]^2}}}\left[\sin\pi(2n-1)x\right.\nonumber\\
\label{AsymmetricFunctions1_2n1}
&-&\left.\frac{1}{\pi(2n-1)\Lambda}\cos\pi(2n-1)x\right]\\
\Psi_{A_{2n}}(\Lambda;x)&=&\sqrt{\frac{2}{1+(2\pi n\Lambda)^2}}\left(\sin2\pi nx\right.\nonumber\\
\label{AsymmetricFunctions1_2n}
&+&\left.2\pi n\Lambda\cos2\pi nx\right).
\end{eqnarray}
\end{subequations}
Their asymptotes are:

for $\Lambda\gg1$:
\begin{subequations}\label{AsymmetricFunctions2}
\begin{eqnarray}\label{AsymmetricFunctions2_0}
\Psi_{A_0}(\Lambda;x)&=&1+\frac{x}{\Lambda}+\left(\frac{x^2}{2}-\frac{1}{12}\right)\frac{1}{\Lambda^2}\\
\Psi_{A_{2n-1}}(\Lambda;x)&=&\left[1-\frac{1}{2}\frac{1}{\pi^2(2n-1)^2}\frac{1}{\Lambda^2}\right]2^{1/2}\!\sin\pi(2n-1)x\nonumber\\
\label{AsymmetricFunctions2_2n1}
&-&\frac{1}{\pi(2n-1)\Lambda}2^{1/2}\cos\pi(2n-1)x\\
\Psi_{A_{2n}}(\Lambda;x)&=&\!\!\left(\!1-\frac{1}{8\pi^2n^2\Lambda^2}\right)2^{1/2}\!\cos2\pi nx\nonumber\\
\label{AsymmetricFunctions2_2n}
&+&\frac{1}{2\pi n\Lambda}2^{1/2}\!\sin2\pi nx;
\end{eqnarray}
\end{subequations}
\begin{figure}
\centering
\includegraphics[width=0.75\columnwidth]{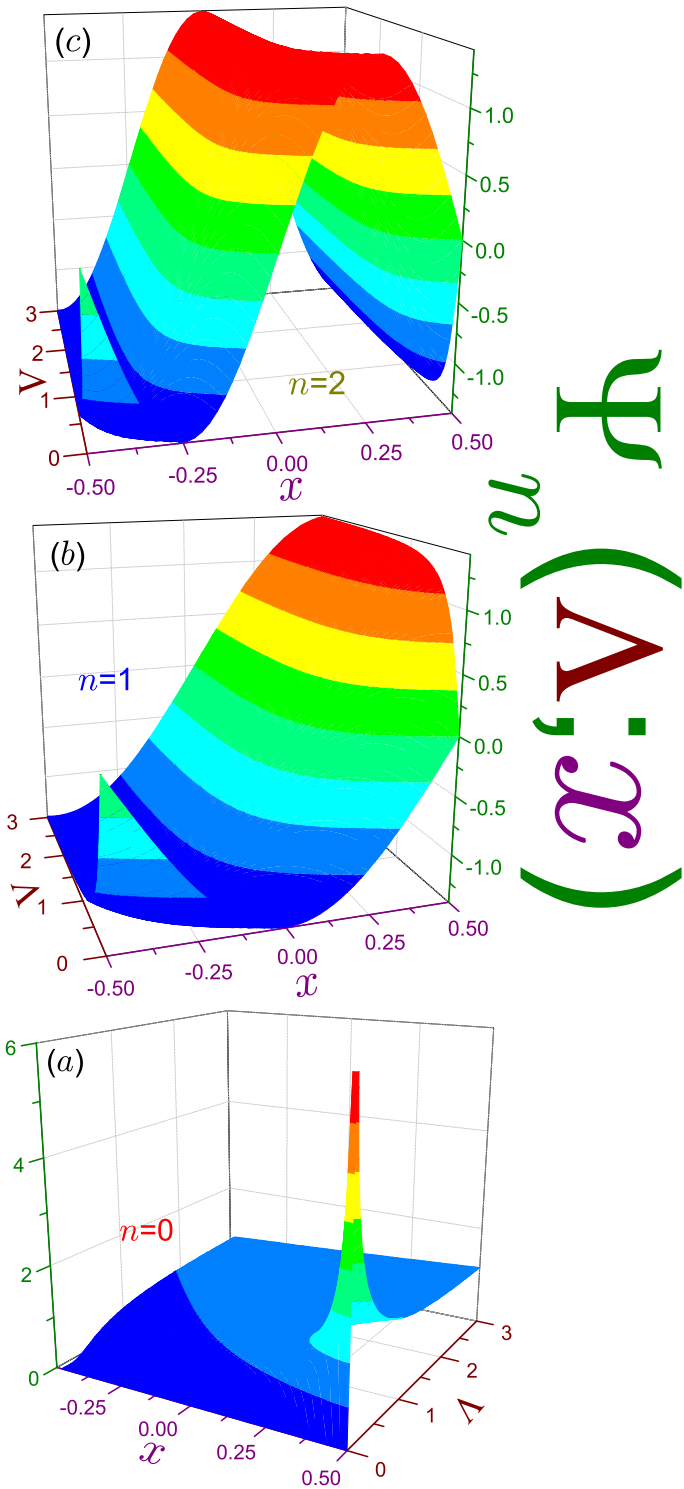}
\caption{\label{FunctionsXasymmetricFig1}
Position wave functions $\Psi_n(\Lambda;x)$ of the asymmetric QW in terms of the coordinate $x$ and extrapolation length $\Lambda$ for (a) the lowest level, $n=0$, (b) first excited orbital, $n=1$, and (c) second excited state, $n=2$. Note different vertical range in panel (a) as compared to its (b) and (c) counterparts.}
\end{figure}

at $\Lambda\ll1$:
\begin{subequations}\label{AsymmetricFunctions3}
\begin{eqnarray}\label{AsymmetricFunctions3_0}
\Psi_{A_0}(\Lambda;x)&=&\left(\frac{2}{\Lambda}\right)^{1/2}\exp\!\left(\frac{x-1/2}{\Lambda}\right)\\
\Psi_{A_{2n-1}}(\Lambda;x)&=&\left[-1+\frac{\pi^2(2n-1)^2}{2}\Lambda^2\right]2^{1/2}\!\cos\pi(2n\!-\!1)x\nonumber\\
\label{AsymmetricFunctions3_2n1}
&+&\Lambda\pi(2n-1)2^{1/2}\!\sin\pi(2n\!-\!1)x\\
\Psi_{A_{2n}}(\Lambda;x)&=&(1-2\pi^2n^2\Lambda^2)\,2^{1/2}\!\sin2\pi nx\nonumber\\
\label{AsymmetricFunctions3_2n}
&+&\Lambda2\pi n2^{1/2}\cos2\pi nx.
\end{eqnarray}
\end{subequations}
They show that at the large extrapolation length each excited Neumann waveform $\Psi_{n+1}^N(x)\equiv\Psi_{A_{n+1}}(\infty;x)$,
\begin{subequations}\label{DirNeuFunctions1}
\begin{align}\label{NeumannFunctions1}
\Psi_{n+1}^N(x)=2^{1/2}\cos\pi(n+1)\left(x-\frac{1}{2}\right),
\intertext{$n=0,1,2,\ldots$, is distorted by the small, proportional to $1/\Lambda$, admixture of the Dirichlet component $\Psi_n^D(x)\equiv\Psi_{A_n}(0;x)$,}
\label{DirichletFunctions1}
\Psi_n^D(x)=2^{1/2}\sin\pi(n+1)\left(x-\frac{1}{2}\right),
\intertext{with the opposite symmetry and with its quantum number $n$ being smaller by one. Relative contribution of the former (latter) dependence decreases (increases) as the Robin distance shrinks, and at the very small de Gennes parameter the overwhelming portion of the total wave function is described by the Dirichlet orbital, which is slightly perturbed by the Neumann counterpart. A transformation of the waveforms from the Neumann BC to the Dirichlet requirement, which is shown in panels (b) and (c) of Fig.~\ref{FunctionsXasymmetricFig1} for the two lowest excited levels, occurs in such a way that the associated energies from Eq.~\eqref{AsymmetricSpectrum1} do not change with the variation of the extrapolation length. On the other hand, ground level energy and function $\Psi_{A_0}$ both are strongly $\Lambda$ dependent quantities; for example, in the Neumann limit the former turns to zero, $\left.E_0\right|_{\Lambda=\infty}=0$, and the latter degenerates into the position independent unity,}
\label{NeumannFunctions0}
\Psi_0^N(x)=1,
\end{align}
\end{subequations}
which, for the large finite Robin length is slightly disturbed by the linear in $x$ admixture, as it follows from Eq.~\eqref{AsymmetricFunctions2_0}. Further decrease of the de Gennes parameter leads to the stronger deviation of the waveform  from uniformity with its gradual accumulation near the wall with the negative extrapolation length until in the limit of extremely small positive $\Lambda$ the particle is firmly affixed to the right surface with the near zero probability to find it away from the interface, as Eq.~\eqref{AsymmetricFunctions3_0} and lower panel of Fig.~\ref{FunctionsXasymmetricFig1} exemplify.

\begin{figure}
\centering
\includegraphics[width=\columnwidth]{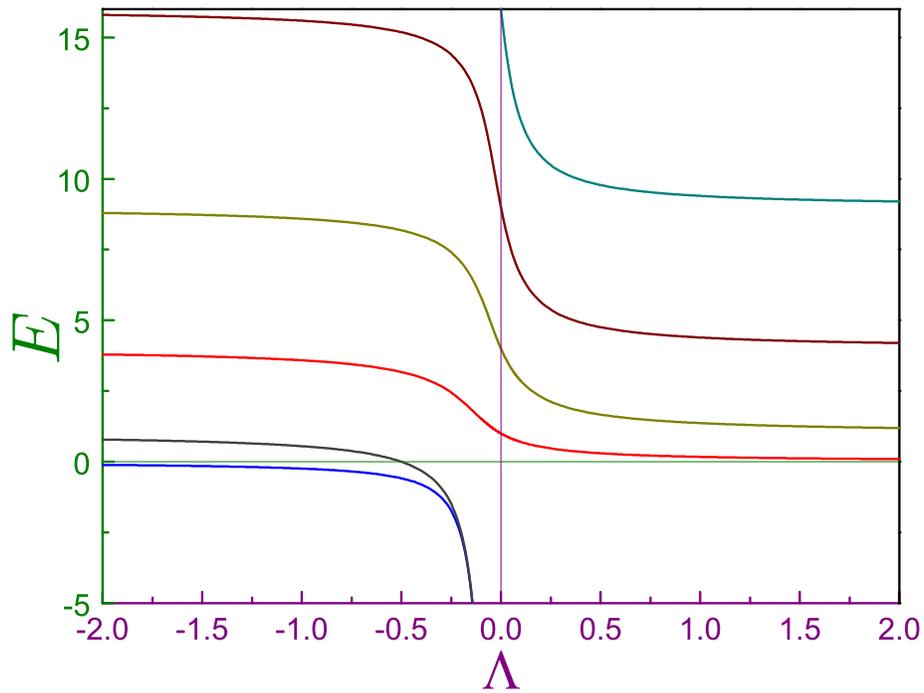}
\caption{\label{EnergiesFig1}
Energies $E_n$ of the symmetric QW as functions of the extrapolation length $\Lambda$. Thin horizontal and vertical lines denote zero energy and Dirichlet BC, respectively.}
\end{figure}

For the symmetric geometry with the same value of the extrapolation parameter at each wall, $\Lambda_-=\Lambda_+\equiv\Lambda$,  position wave functions are conveniently separated into those of the symmetric, $\Psi^S(-x)=\Psi^S(x)$, states:
\begin{subequations}\label{FunctionsSymmetric1}
\begin{align}\label{FunctionsSymmetric1_Even1}
\Psi_{S_n}^S(\Lambda;x)&=\left(\frac{2}{1+\frac{\sin\pi\sqrt{E_{S_n}^S}}{\pi\sqrt{E_{S_n}^S}}}\right)^{1/2}\!\!\cos\!\left(\!\pi\sqrt{E_{S_n}^S}\,x\right)\\
\intertext{[for the lowest Neumann level, $n=0$, at $\Lambda=\pm\infty$ it degenerates, as stated above, to the position independent unity, Eq.~\eqref{NeumannFunctions0}], and their antisymmetric, $\Psi^A(-x)=-\Psi^A(x)$, counterparts:}
\label{FunctionsSymmetric1_Odd}
\Psi_{S_n}^A(\Lambda;x)&=\left(\frac{2}{1-\frac{\sin\pi\sqrt{E_{S_n}^A}}{\pi\sqrt{E_{S_n}^A}}}\right)^{1/2}\!\!\sin\!\left(\!\pi\sqrt{E_{S_n}^A}\,x\right),
\end{align}
\end{subequations}
where the BC dependent energies $E_{S_n}^{S,A}(\Lambda)$ are found from
\begin{subequations}\label{EnergySymmetric1}
\begin{eqnarray}\label{EnergySymmetric1_Even}
\Lambda\pi\sqrt{E_{S_n}^S}=\cot\!\frac{\pi\sqrt{E_{S_n}^S}}{2}\\
\label{EnergySymmetric1_Odd}
-\Lambda\pi\sqrt{E_{S_n}^A}=\tan\!\frac{\pi\sqrt{E_{S_n}^A}}{2},
\end{eqnarray}
\end{subequations}
$n=0,1,2,\ldots$. Above equations are easily modified for the negative energies that are characteristic for the lowest even state at the arbitrary negative de Gennes distance and for its neighboring odd fellow at $-1/2<\Lambda<-0$ \cite{AlHashimi1,Olendski1}. In the limits of the extremely small and large Robin lengths, Eqs.~\eqref{EnergySymmetric1} yield:
\begin{subequations}\label{EnergySymmetricLimit1}
\begin{align}\label{EnergySymmetricLimit1_Small}
E_{S_n}(\Lambda)=&(n+1)^2(1-4\Lambda),\quad|\Lambda|\ll1\\
\label{EnergySymmetricLimit1_Large}
E_{S_n}(\Lambda)=&\left\{\begin{array}{cl}
\frac{2}{\pi^2}\frac{1}{\Lambda}\left(1-\frac{4}{\Lambda}\right),&n=0\\
n^2+\frac{4}{\pi^2}\frac{1}{\Lambda}-\frac{4}{\pi^4n^2}\frac{1}{\Lambda^2},&n\geq1,\\
\end{array}\right\},\quad\frac{1}{|\Lambda|}\ll1,\\
\intertext{where the absence of the superscript at the energy indicates that these dependencies are valid both for the symmetric and antisymmetric orbitals. Note that in Eq.~\eqref{EnergySymmetricLimit1_Small} only the terms up to the linear power of the extrapolation length are retained whereas the expansion from Eq.~\eqref{EnergySymmetricLimit1_Large} takes into account the items with up to the quadratic dependence on $|\Lambda|^{-1}$. The reason for this will become clear in subsec.~\ref{SubSec_AnalyticSymmetric}. The lowest even ($e$) and odd ($o$) levels whose energies at $\Lambda\rightarrow-\infty$ are described by Eq.~\eqref{EnergySymmetricLimit1_Large} with $n=0$ and $n=1$, are splitting off with the decrease of the absolute value of the negative $\Lambda$ from their positive counterparts with their negative energies unrestrictedly falling and simultaneously approaching each other as the Robin length tends to zero from the left:}
\label{EnergySymmetricLimit1_MinusZero}
E_{\left\{_e^o\right\}}(\Lambda)=&-\frac{1}{\pi^2\Lambda^2}\left(1\mp4e^{-|\Lambda|^{-1}}\right),\quad\Lambda\rightarrow-0.
\intertext{Note that at $\Lambda<0$ the energy of the lowest even state is always negative while its higher lying odd neighbor leaves the positive part of the spectrum at $\Lambda=-1/2$ \cite{Olendski1} around which point its energy is governed by the following dependence:}
\label{EnergySymmetricLimit1_MinusOneHalf}
E_o(\Lambda)=-&\frac{24}{\pi^2}\!\left(\Lambda+\frac{1}{2}\right)+\frac{72}{5\pi^2}\left(\Lambda+\frac{1}{2}\right)^2+\dots,\quad\Lambda+\frac{1}{2}\rightarrow0.
\end{align}
\end{subequations}
Corresponding position waveforms read:
\begin{subequations}\label{FunctionsSymmetricLimit1}
\begin{eqnarray}
\Psi_{S_n}^{\left\{_S^A\right\}}(\Lambda;x)=2^{1/2}(1-\Lambda)\nonumber\\
\label{FunctionsSymmetricLimit1_Small}
\times\left\{\begin{array}{c}
\sin2n\pi(1-2\Lambda)x,\,n=1,2,\ldots\\\cos(2n+1)\pi(1-2\Lambda)x,\,n=0,1,\ldots
\end{array}\!\!\!\!\right\},\,|\Lambda|\ll1\\
\Psi_{S_n}^{\left\{_S^A\!\right\}}\!(\Lambda;x)=\nonumber\\
\left\{\!\!\!\begin{array}{c}
\begin{array}{l}
2^{1/2}\!\left[1-\frac{1}{\pi^2(2n+1)^2\Lambda}+\frac{11}{2}\frac{1}{\pi^4(2n+1)^4\Lambda^2}\right]\\
\times\sin\!\left(2n\pi+\pi+\frac{2}{\pi(2n+1)\Lambda}-\frac{4}{\pi^3(2n+1)^3\Lambda^2}\!\right)\!x
\end{array}\\
\left\{\!\!\begin{array}{cl}
1\!+\!\frac{1}{\Lambda}\!\left(\!-x^2\!+\!\frac{1}{12}\right)\!+\frac{1}{\Lambda^2}\!\left(\frac{x^4+x^2}{6}-\frac{3}{160}\right),&\!\!n=0\\
\left\{\begin{array}{l}
2^{1/2}\!\!\left[1-\frac{1}{\pi^2(2n)^2\Lambda}+\frac{11}{2}\frac{1}{\pi^4(2n)^4\Lambda^2}\right]\\
\times\cos\left(2n\pi+\frac{1}{\pi n\Lambda}-\frac{4}{\pi^3(2n)^3\Lambda^2}\right)\!x
\end{array}\right\},
&\!\!n\geq1
\end{array}
\right.
\end{array}\!\!\!\!\!\!\right\},\nonumber\\
\label{FunctionsSymmetricLimit1_Large}
\frac{1}{|\Lambda|}\!\ll\!1\\
\Psi_{\left\{_e^o\right\}}(\Lambda;x)=\frac{1}{|\Lambda|^{1/2}}\left\{\!\!\begin{array}{c}
\left\{\begin{array}{cc}
2\frac{x}{|\Lambda|}\,e^{-\frac{1}{2|\Lambda|}},&|x|\ll|\Lambda|\\
\pm e^{\frac{1}{|\Lambda|}\left(|x|-1/2\right)},&{\rm otherwise}
\end{array}\right\}\\
e^{\frac{1}{|\Lambda|}\left(|x|-1/2\right)}
\end{array}
\!\!\right\},\nonumber\\
\label{FunctionsSymmetricLimit1_MinusZero}
\Lambda\rightarrow-0,\\
\Psi_o(\Lambda;x)=2\sqrt{3}\left[x-\left(\frac{3}{5}\,x-4x^3\right)\left(\Lambda+\frac{1}{2}\right)\right],\nonumber\\
\label{FunctionsSymmetricLimit1_MinusOneHalf}
\Lambda+\frac{1}{2}\rightarrow0,
\end{eqnarray}
\end{subequations}
where the plus or minus sign of the odd split-off level in Eq.~\eqref{FunctionsSymmetricLimit1_MinusZero} corresponds to that of $x$. Observe that the functions from Eqs.~\eqref{FunctionsSymmetricLimit1} up to and including the largest power of nonvanishing small perturbation term in them do satisfy the normalization condition, Eq.~\eqref{Normalization2}.

\begin{figure}
\centering
\includegraphics[width=0.75\columnwidth]{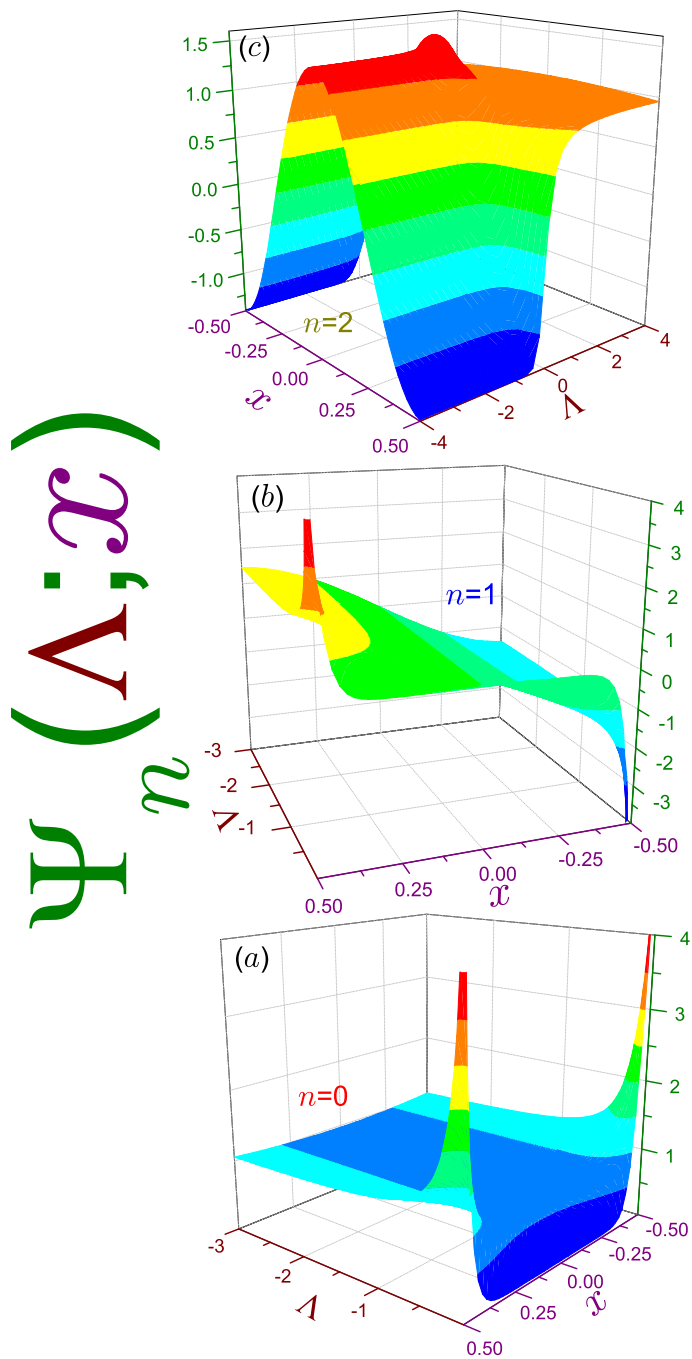}
\caption{\label{FunctionsXsymmetricFig1}
Position wave functions $\Psi_n(\Lambda;x)$ of the symmetric geometry in terms of the coordinate $x$ and extrapolation length $\Lambda$ for (a) the lowest level at the negative Robin distances, (b) first excited orbital at $\Lambda<0$ and (c) the state that at the large positive $\Lambda$ possesses the lowest energy. Note different vertical ranges in each figure and distinct $\Lambda$ interval in panel (c) as compared to its (a) and (b) counterparts. To better exhibit the characteristic features of each dependence, different viewing perspectives are used in the subplots.}
\end{figure}

Even though the shape of the spectrum of the symmetric QW is well known \cite{AlHashimi1,Slachmuylders1,Olendski1,Cacciapuoti1}, for completeness we plot it in Fig.~\ref{EnergiesFig1}. It shows that at the large negative Robin lengths  the energy $E_{S_n}$  with $n\geq2$ monotonically decreases with the growth of the de Gennes distance according to Eq.~\eqref{EnergySymmetricLimit1_Large} and at $\Lambda=0$ the orbital transforms into the $(n-2)$th Dirichlet level with $E_n(\Lambda)$ in its vicinity given by Eq.~\eqref{EnergySymmetricLimit1_Small}. Further growth of the positive length decreases more the corresponding energy and at $1/\Lambda\ll1$ each state approaches asymptotically the $(n-2)$th Neumann level. The two lowest at $\Lambda=-\infty$ energies $E_{e,o}$ decline with the decrease of the absolute value of the Robin length and close to the point of non-analyticity $\Lambda=0$ they tend to the unrestrictedly large negative values according to Eq.~\eqref{EnergySymmetricLimit1_MinusZero} whereas the corresponding position waveforms $\Psi_n(\Lambda;x)$ are symmetrically (for $n=0$) or anti symmetrically ($n=1$) localized at the surfaces $x=\pm1/2$, as panels (a) and (b) of Fig.~\ref{FunctionsXsymmetricFig1} exemplify. Subplot (c) demonstrates an evolution of the $n=2$ Neumann function $\Psi(x)$ to its zeroth counterpart as the extrapolation length scans the $\Lambda$ axis from its large negative to huge  positive values. Note that its global maximum $\Psi_{max}$, which for the even $n$ is always located in the middle of the well, $x=0$, is equal to $1.598$ and is achieved at $\Lambda=-0.18$. 

\section{Symmetric QW, $\Lambda_-=\Lambda_+=\Lambda$}\label{Sec_Symmetric1}
This chapter is devoted to the analysis of quantum-information measures of the Robin QW with the same extrapolation lengths on both interfaces and the next section describes them for the antisymmetric geometry. Since it does not cause any confusion, in the discussion below the subscripts '$S$' or '$A$' that were used before are dropped.
\subsection{Analytic results}\label{SubSec_AnalyticSymmetric}
With the help of dependencies from Eqs.~\eqref{FunctionsSymmetric1}, position quantum information measures can be analytically calculated as functions of the energy $E_n(\Lambda)$ but since the expression for the Shannon entropy is quite unwieldy, below only the formulae of the Fisher information and Onicescu energy are provided:
\begin{subequations}\label{FisherOnicescuX_Anal1}
\begin{eqnarray}\label{FisherX_Anal1}
I_{x_n}^{\left\{_S^A\!\right\}}(\Lambda)&=&4\pi^2\frac{1\pm\frac{\sin\pi\sqrt{E_n}}{\pi\sqrt{E_n}}}{1\mp\frac{\sin\pi\sqrt{E_n}}{\pi\sqrt{E_n}}}\,E_n\\
O_{x_n}^{\left\{_S^A\!\right\}}(\Lambda)&=&\frac{1}{2}\frac{1}{\left(1\mp\frac{\sin\pi\sqrt{E_n}}{\pi\sqrt{E_n}}\right)^2}\nonumber\\
\label{OnicescuX_Anal1}
&\times&\left[3+\frac{\sin\pi\sqrt{E_n}}{\pi\sqrt{E_n}}\left(\cos\pi\sqrt{E_n}\mp4\right)\right].
\end{eqnarray}
\end{subequations}
Similar to the wave functions, they remain valid for the negative energies too. Their asymptotic expressions are calculated as:
\begin{subequations}\label{EntropyXlimits1}
\begin{eqnarray}\label{EntropyXlimits1_Small}
S_{x_n}(\Lambda)&=&\ln2-1+2\Lambda,\quad|\Lambda|\ll1,\quad n=0,1,\ldots\\
\label{EntropyXlimits1_Large}
S_{x_n}(\Lambda)&=&\left\{
\begin{array}{cc}
-\frac{1}{90}\frac{1}{\Lambda^2},&n=0\\
\ln2-1+2\frac{3-4\ln2}{\pi^2n^2}\frac{1}{\Lambda},&n\geq1
\end{array}\right\},\quad\frac{1}{|\Lambda|}\ll1\\
\label{EntropyXlimits1_MinusZero}
S_{x_{e,o}}(\Lambda)&=&\ln|\Lambda|+1,\quad\Lambda\rightarrow-0\\
\label{EntropyXlimits1_MinusOneHalf}
S_{x_o}(\Lambda)&=&\frac{2}{3}-\ln3-\frac{8}{25}\left(\Lambda+\frac{1}{2}\right),\quad\Lambda+\frac{1}{2}\rightarrow0
\end{eqnarray}
\end{subequations}
\begin{subequations}\label{FisherXlimits1}
\begin{eqnarray}\label{FisherXlimits1_Small}
I_{x_n}(\Lambda)&=&4\pi^2(n+1)^2(1-8\Lambda),\quad|\Lambda|\ll1,\quad n=0,1,\ldots\\
\label{FisherXlimits1_Large}
I_{x_n}(\Lambda)&=&\left\{\begin{array}{cc}
\frac{4}{3}\frac{1}{\Lambda^2},&n=0\\
4\pi^2n^2+\frac{16}{\pi^2n^2}\frac{1}{\Lambda^2},&n\geq1
\end{array}\right\},\quad\frac{1}{|\Lambda|}\ll1\\
\label{FisherXlimits1_MinusZero}
I_{x_{e,o}}(\Lambda)&=&\frac{4}{|\Lambda|^2},\quad\Lambda\rightarrow-0\\
\label{FisherXlimits1_MinusOneHalf}
I_{x_o}(\Lambda)&=&48+\frac{192}{5}\left(\Lambda+\frac{1}{2}\right),\quad\Lambda+\frac{1}{2}\rightarrow0
\end{eqnarray}
\end{subequations}
\begin{subequations}\label{OnicescuXlimits1}
\begin{eqnarray}\label{OnicescuXlimits1_Small}
O_{x_n}(\Lambda)&=&\frac{3}{2}-3\Lambda,\quad|\Lambda|\ll1,\quad n=0,1,\ldots\\
\label{OnicescuXlimits1_Large}
O_{x_n}(\Lambda)&=&
\left\{\begin{array}{cc}
1+\frac{1}{45}\frac{1}{\Lambda^2},&n=0\\
\frac{3}{2}-\frac{1}{\pi^2n^2}\frac{1}{\Lambda},&n\geq1
\end{array}\right\},\quad\frac{1}{|\Lambda|}\ll1\\
\label{OnicescuXlimits1_MinusZero}
O_{x_{e,o}}(\Lambda)&=&\frac{1}{2|\Lambda|},\quad\Lambda\rightarrow-0\\
\label{OnicescuXlimits1_MinusOneHalf}
O_{x_o}(\Lambda)&=&\frac{9}{5}+\frac{144}{175}\left(\Lambda+\frac{1}{2}\right),\quad\Lambda+\frac{1}{2}\rightarrow0.
\end{eqnarray}
\end{subequations}
\begin{subequations}\label{CGLXlimits1}
\begin{eqnarray}\label{CGLXlimits1_Small}
CGL_{x_n}(\Lambda)&=&\frac{3}{e},\quad|\Lambda|\ll1,\quad n=0,1,\ldots\\
\label{CGLXlimits1_Large}
CGL_{x_n}(\Lambda)&=&
\left\{\begin{array}{cc}
1+\frac{1}{90}\frac{1}{\Lambda^2},&n=0\\
\frac{3}{e}-\frac{8}{e}\frac{3\ln2-2}{\pi^2n^2}\frac{1}{\Lambda},&n\geq1
\end{array}\right\},\quad\frac{1}{|\Lambda|}\ll1\\
\label{CGLXlimits1_MinusZero}
CGL_{x_{e,o}}(\Lambda)&=&\frac{e}{2},\quad\Lambda\rightarrow-0\\
\label{CGLXlimits1_MinusOneHalf}
CGL_{x_o}(\Lambda)&=&e^{2/3}\left[\frac{3}{5}+\frac{72}{175}\left(\Lambda+\frac{1}{2}\right)\right],\quad\Lambda+\frac{1}{2}\rightarrow0.
\end{eqnarray}
\end{subequations}
Note that the tiny deviation from the Neumann structure results in the quadratic dependence of the position Fisher information $I_{x_n}(\Lambda)$ on the small perturbation $\Lambda^{-1}$. The same is true for the Shannon entropy and Onicescu energy of the lowest level whereas all other measures are the linear functions of the slight distortion of both the Dirichlet and Neumann QWs. This is the reason why we keep in Eqs.~\eqref{EnergySymmetricLimit1_Small} and \eqref{FunctionsSymmetricLimit1_Small}, on the one hand, and, to the contrary, Eqs.~\eqref{EnergySymmetricLimit1_Large} and \eqref{FunctionsSymmetricLimit1_Large}, different powers of the corresponding disturbances $\Lambda$ and $\Lambda^{-1}$. It has to be also pointed out that Fisher informations from Eq.~\eqref{FisherXlimits1_MinusZero} do coincide with the corresponding expression for the single Robin wall \cite{Olendski3} whereas the disequilibria, Eq.~\eqref{OnicescuXlimits1_MinusZero}, are just one half of that of the lonely attractive surface \cite{Olendski3}.

Momentum waveforms $\Phi_n(\Lambda;k)$ are written as:
\begin{subequations}\label{FunctionsMomentumSymmetric1}
\begin{eqnarray}
\Phi_n^S(\Lambda;k)&=&\frac{2}{\pi^{1/2}\left(1+\frac{\sin\pi\sqrt{E_n^S}}{\pi\sqrt{E_n^S}}\right)^{1/2}}\nonumber\\
\label{FunctionsMomentumSymmetric1_Even1}
&\times&\frac{k\sin\frac{k}{2}\cos\frac{\pi\sqrt{E_n^S}}{2}-\pi\sqrt{E_n^S}\cos\frac{k}{2}\sin\frac{\pi\sqrt{E_n^S}}{2}}{k^2-\pi^2E_n^S}\\
\Phi_n^A(\Lambda;k)&=&\frac{2}{\pi^{1/2}\left(1-\frac{\sin\pi\sqrt{E_n^A}}{\pi\sqrt{E_n^A}}\right)^{1/2}}\nonumber\\
\label{FunctionsMomentumSymmetric1_Odd1}
&\times&\frac{k\cos\frac{k}{2}\sin\frac{\pi\sqrt{E_n^A}}{2}-\pi\sqrt{E_n^A}\sin\frac{k}{2}\cos\frac{\pi\sqrt{E_n^A}}{2}}{k^2-\pi^2E_n^A},
\end{eqnarray}
\end{subequations}
$n=0,1,\ldots$. Observe that for the symmetric BCs these even and odd functions of the momentum $k$ are real. Then, for calculating corresponding Fisher informations $I_{k_n}(\Lambda)$, it is convenient to use a method applied before for the Dirichlet well \cite{LopezRosa1}; namely, in this case the integrand in Eq.~\eqref{Fisher1_K} becomes $4\left[\Phi_n'(k)\right]^2$, and using the reciprocity between position and momentum spaces, one replaces infinite $k$ integration by the finite $x$ one:
\begin{equation}\label{FisherK2}
I_{k_n}(\Lambda)=4\int_{-1/2}^{1/2}x^2\Psi_n^2(\Lambda;x)dx.
\end{equation}
For functions from Eqs.~\eqref{FunctionsSymmetric1}, the calculation of the mean value of the square of the position yields:
\begin{eqnarray}
I_{k_n}^{\left\{_S^A\right\}}(\Lambda)=\frac{1}{3}\frac{1}{1\mp\frac{\sin\pi\sqrt{E_n}}{\pi\sqrt{E_n}}}\nonumber\\
\label{FisherK3}
\times\left[1\mp\frac{6\pi\sqrt{E_n}\cos\pi\sqrt{E_n}+3(\pi^2E_n-2)\sin\pi\sqrt{E_n}}{\pi^3\sqrt{E_n^3}}\right];
\end{eqnarray}
in particular:
\begin{subequations}\label{FisherK4}
\begin{eqnarray}
I_{k_n}(\Lambda)&=&\frac{1}{3}\left[1-\frac{6}{\pi^2(n+1)^2}\right](1+4\Lambda),\nonumber\\
\label{FisherK4_Dir1}
n&=&0,1,2,\ldots,\quad|\Lambda|\ll1\\
I_{k_n}(\Lambda)&=&\left\{
\begin{array}{cc}
\frac{1}{3}-\frac{2}{45}\frac{1}{\Lambda},&n=0\\
\frac{1}{3}\left(1+\frac{6}{\pi^2n^2}\right)+\frac{4}{3}\frac{\pi^2n^2-12}{\pi^4n^4}\frac{1}{\Lambda},&n\geq1
\end{array}
\right\},\nonumber\\
\label{FisherK4_Neu1}
&&\frac{1}{|\Lambda|}\ll1.
\end{eqnarray}
\end{subequations}
Then, the product of the position and momentum Fisher informations in the close vicinity of the Dirichlet or Neumann BC is, respectively:
\begin{subequations}\label{FisherK7}
\begin{eqnarray}
I_{x_n}(\Lambda)I_{k_n}(\Lambda)&=&\frac{4}{3}\left[\pi^2(n+1)^2\!-\!6\right](1\!-\!4\Lambda),\nonumber\\
\label{FisherK7_Dir1}
n&=&0,1,2,\ldots,\quad|\Lambda|\ll1\\
I_{x_n}(\Lambda)I_{k_n}(\Lambda)&=&\!\!\!\left\{
\begin{array}{cc}
\frac{2}{9}\frac{1}{\Lambda^2},&n=0\\
\frac{4}{3}\!\left(\pi^2n^2\!+\!6\right)\!+\!\frac{16}{3}\!\left(1\!-\!\frac{12}{\pi^2n^2}\right)\frac{1}{\Lambda},&n\geq1
\end{array}
\!\!\!\right\},\nonumber\\
\label{FisherK7_Neu1}
&&\frac{1}{|\Lambda|}\ll1.
\end{eqnarray}
\end{subequations}
Note that at $\Lambda=0$ Eq.~\eqref{FisherK7_Dir1} coincides, as expected, with the Dirichlet result obtained recently \cite{LopezRosa1,Saha1}. It is also important to stress that the ground level does violate Eq.~\eqref{FisherUncertainty1} in the vicinity of the Neumann BC. Some other analytic expressions will be derived and analyzed during the discussion of the corresponding momentum measures.

\subsection{Momentum densities}\label{SubSec_EnergySymmetric}

\begin{figure}
\centering
\includegraphics[width=0.8\columnwidth]{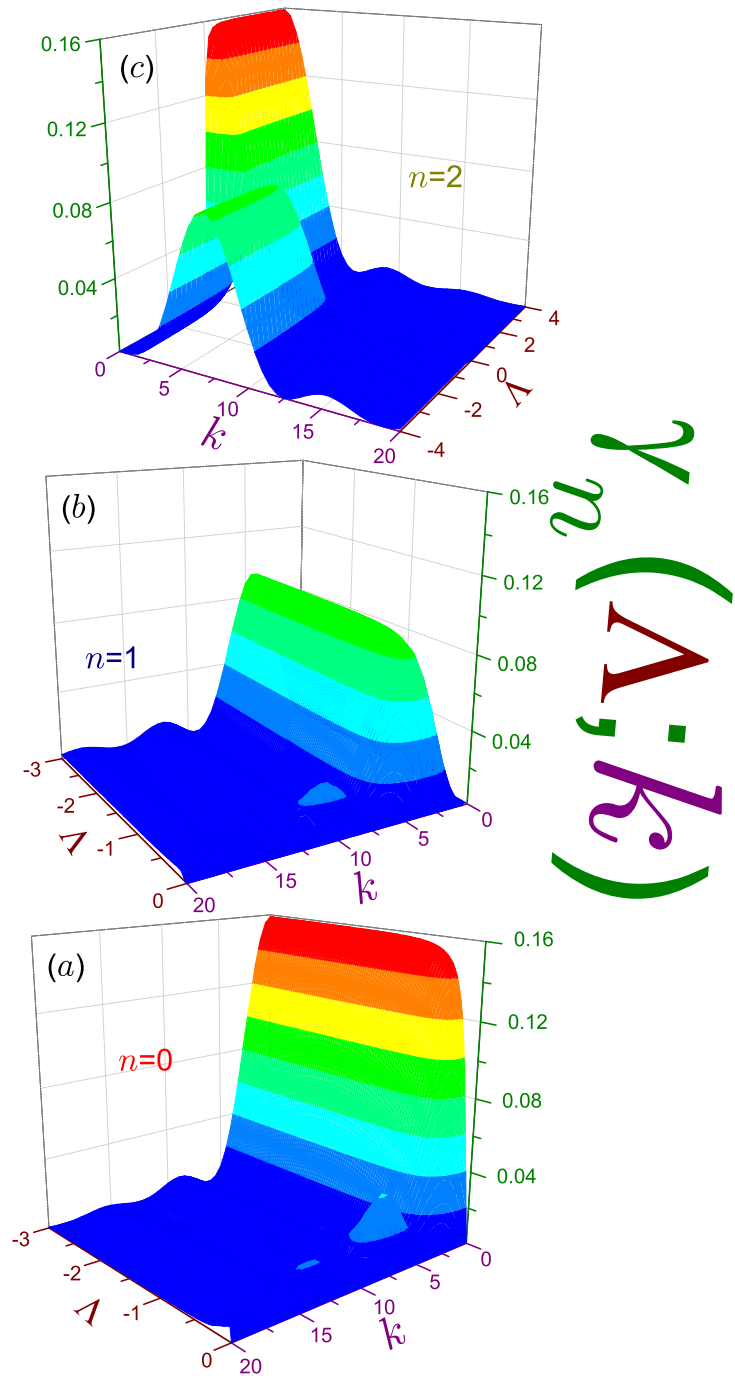}
\caption{\label{DensityKsymmetricFig1}
Momentum densities $\gamma_n(\Lambda;k)$ of the symmetric geometry as  functions of the momentum $k$ and extrapolation length $\Lambda$. Each panel has its position function counterpart from Fig.~\ref{FunctionsXsymmetricFig1}. Since $\gamma_n(\Lambda;k)$ is an even function of its argument $k$, $\gamma_n(\Lambda;-k)=\gamma_n(\Lambda;k)$, only the dependencies for positive momenta are shown.}
\end{figure}

Momentum densities, which in the limit of $\Lambda=0$ or $\Lambda=\infty$ degenerate to the Dirichlet \cite{Robinett1,Majernik1,Majernik2,Majernik3,Olendski2} or Neumann \cite{Olendski2} dependencies, respectively, are depicted in Fig.~\ref{DensityKsymmetricFig1} where the upper panel shows an evolution  of $\gamma_n({\Lambda;k})$ as the Robin length sweeps the whole $\Lambda$ axis.  It is known that the global maximum of the Neumann momentum density for the quantum index $n\ge1$ is achieved at the nonzero momentum and is smaller than its $n=0$ counterpart of the magnitude of $1/(2\pi)=0.1592$ located at $k=0$ \cite{Olendski2}. As is clearly seen in panel (c), a transformation from $n=2$ Neumann, $\Lambda=-\infty$, waveform with zero minimum at $k=0$ to that with $n=0$ for $\Lambda\rightarrow+\infty$ takes place through the lowest, $n=0$, Dirichlet, $\Lambda=0$, orbital whose global zero-momentum extremum is $4/\pi^3=0.1290$ \cite{Olendski2}. Panels (a) and (b) of Fig.~\ref{DensityKsymmetricFig1} show that a characteristic feature of the densities $\gamma$ of the two lowest split-off levels that exist for the negative extrapolation lengths only is their decrease as the Robin distance approaches zero from the left and in its very vicinity they fade as:
\begin{equation}\label{MomentumDensityAtZero1}
\gamma_{e,o}(\Lambda;k)=\frac{2}{\pi}|\Lambda|\left(\frac{\cos\frac{k}{2}+|\Lambda|k\sin\frac{k}{2}}{1+|\Lambda|^2k^2}\right)^2,\quad\Lambda\rightarrow-0.
\end{equation}
Note that at any $n$, $\Lambda$ and $k$ the momentum density is smaller than unity:
\begin{equation}
\gamma_n(\Lambda;k)<1,
\end{equation}
what means that the corresponding entropies $S_{k_n}(\Lambda)$ from Eq.~\eqref{Entropy1_K} are always positive.

\begin{figure}
\centering
\includegraphics[width=\columnwidth]{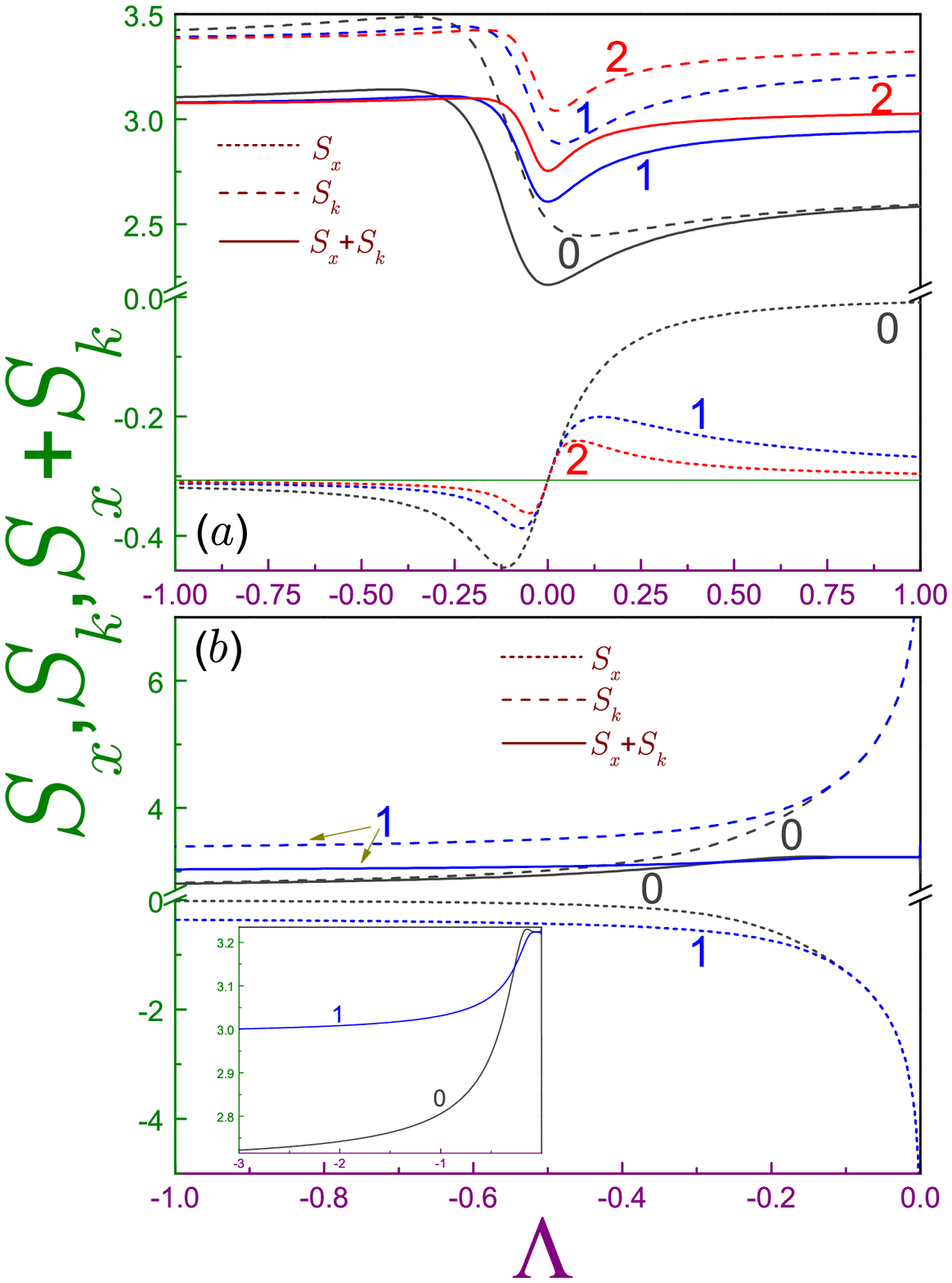}
\caption{\label{EntropySymmetricFig1}
Position $S_x$ (dotted lines), momentum $S_k$ (dashed curves) quantum information entropies and their sums $S_x+S_k$ (solid shapes) of the symmetric geometry versus extrapolation length $\Lambda$ for (a) three lowest at the large Robin lengths levels and (b) two lowest at the large negative de Gennes distance states. Numbers near the curves denote corresponding quantum indexes $n$. Note vertical line break from $0$ to $2.2$ in panel (a) and from $0$ to $2.7$ in panel (b). Thin horizontal line in the upper subplot designates $\ln2-1\approx-0.30685$ that is a position entropy for all Dirichlet and excited Neumann states \cite{Olendski2}. Inset in window (b) shows the vertically enlarged view of the sums of the two entropies.}
\end{figure}
\subsection{Shannon entropy}\label{SubSec_Shannon}
\begin{figure}
\centering
\includegraphics[width=\columnwidth]{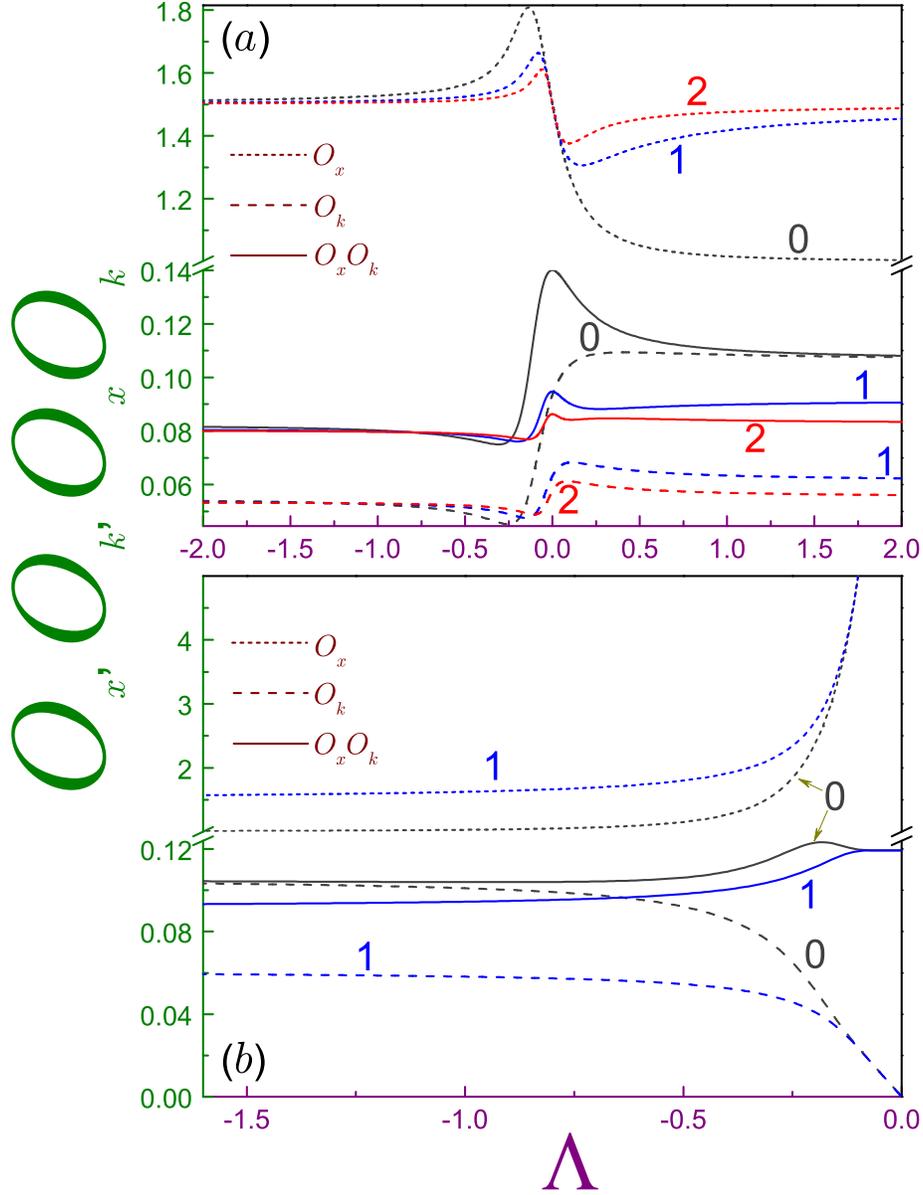}
\caption{\label{OnicescuSymmetricFig1}
Onicescu position $O_x$ (dotted lines), momentum $O_k$ (dashed curves) energies and their products $O_xO_k$ (solid lines) of the symmetric geometry versus extrapolation length $\Lambda$ for (a) three lowest at the large Robin lengths levels and (b) two lowest at the large negative de Gennes distance states. Note vertical line break from $0.141$ to $1.002$ in panel (a) and from $0.124$ to $1$ in panel (b).}
\end{figure}

\begin{figure}
\centering
\includegraphics[width=\columnwidth]{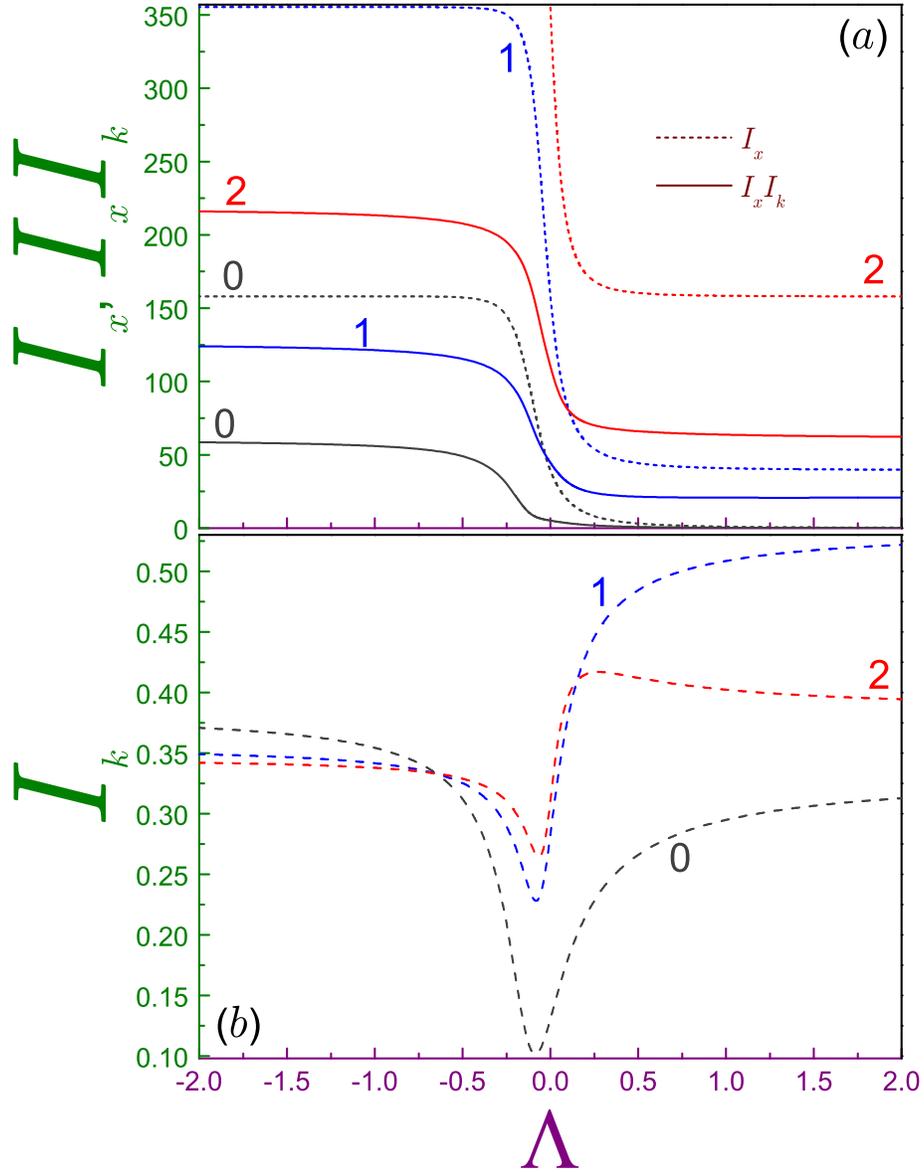}
\caption{\label{FisherSymmetricFig1}
a) Position (dotted lines) Fisher information $I_{x_n}$ and its product with momentum counterpart  $I_{x_n}I_{k_n}$ (solid curves) as functions of the extrapolation length $\Lambda$ for the symmetric QW. b) Momentum Fisher information (dashed lines) of the symmetric QW.}
\end{figure}

The form of the densities $\rho_n(\Lambda;x)$ and $\gamma_n(\Lambda;k)$ defines the corresponding quantum information measures from Eqs~\eqref{Entropy1}, \eqref{Fisher1} - \eqref{CGLdefinition1}. Fig.~\ref{EntropySymmetricFig1} exhibits the evolution of the position $S_x$, momentum $S_k$ Shannon entropies and their sum $S_t$ as a function of the extrapolation length for several low-lying orbitals denoted by the corresponding number near the curve. Here and in subsequent Figs.~\ref{OnicescuSymmetricFig1} - \ref{FisherSymmetricFig2} the following convention is applied: if the $\Lambda$-axis shows only negative Robin distances, the levels are counted in the ascending order their energies possess at $\Lambda=-\infty$ whereas if both positive and negative de Gennes values are used, the counting uses the quantum numbers at $\Lambda=+\infty$. It is seen that position entropies are always negative what means that the parts of the corresponding density, which are larger than unity, in their contribution to the integral from Eq.~\eqref{Entropy1_X} overweigh those with $\rho_n(\Lambda;x)\leq1$ \cite{Olendski2}. It is known that for any combination of the Dirichlet and Neumann BCs the quantum-number-independent entropy $S_{x_n}$ is equal to $\ln2-1\approx-0.30685$ with the only exception being the lowest Neumann state with $S_{x_0}(+\infty)=0$ \cite{Olendski2}. Asymptotic approach to these limiting values according to Eqs.~\eqref{EntropyXlimits1_Small} and~\eqref{EntropyXlimits1_Large} is seen in the figure. Since zero entropy in our choice of units means a complete indetermination of the particle location, the stronger deviation from it corresponds to more information about electron position or momentum. As panel (a) depicts, the position entropy of the lowest at $\Lambda=+\infty$ level reaches its minimum of ${S_{x_0}}_{min}=-0.4534$ at $\Lambda=-0.116$. Note that its location is slightly different from the position of the global maximum of the corresponding waveform specified above, see Fig.~\ref{FunctionsXsymmetricFig1}(c). All other higher-lying states are characterized by the $n$-dependent minimum at the negative Robin distance and maximum at $\Lambda>0$ that are not located symmetrically with respect to the Dirichlet BC with their absolute values decreasing for the increasing quantum index what is a reflection of the fact that the levels with the greater energy are less sensitive to the BC variation. Position entropies of the two split-off states monotonically decrease as the Robin distance approaches zero from the left, as dotted lines in panel (b) show, and at the very small negative extrapolation lengths they diverge as its natural logarithm, according to Eq.~\eqref{EntropyXlimits1_MinusZero}.

Since in known literature \cite{Gradshteyn1,Prudnikov1,Brychkov1} there are no analytic formulae for the integrals in momentum  Shannon entropies with the wave functions from Eqs.~\eqref{FunctionsMomentumSymmetric1}, their direct numerical quadrature was used in calculating the results presented in Fig.~\ref{EntropySymmetricFig1}. It shows that there is an interval of the negative extrapolation lengths where $S_{k_n}$ is not a monotonically increasing function of the index $n$, what was the case for any combination of the Dirichlet and Neumann BCs \cite{Olendski2}. This nonmonotonicity is also inherited by the sum $S_{t_n}$. For each level, the momentum entropy $S_{k_n}$ has $n$-dependent maximum at the negative $\Lambda$ and minimum at the positive Robin distances. Corresponding extremum at $\Lambda<0$ is characteristic also for the sum $S_t$ whereas its global minimum, as our numerical results show, is always located precisely at $\Lambda=0$. This means that the momentum entropy in the near vicinity of the Dirichlet BC is described by the dependence where the linear term is exactly opposite to its position counterpart from Eq.~\eqref{EntropyXlimits1_Small}:
\begin{equation}\label{EntropyKlimits1_Small}
S_{k_n}(\Lambda)=S_{k_n}(0)-2\Lambda,\quad|\Lambda|\ll1.
\end{equation}
As a result, the lowest-order admixture to the total entropy $S_t$ is proportional to the square of the small deviation $\Lambda$
\begin{equation}\label{EntropyTlimits1_Small}
S_{t_n}(\Lambda)=S_{t_n}(0)+c_n\Lambda^2,\quad|\Lambda|\ll1,
\end{equation}
what for the positive index-dependent coefficient $c_n$ creates a global minimum of $S_{x_n}(\Lambda)+S_{k_n}(\Lambda)$ on the length axis. In other words, the Dirichlet BC comes closest to saturating the Bia{\l}ynicki-Birula-Mycielski-Beckner inequality \eqref{EntropicInequality1} with its right-hand side being equal to $2.1447$ whereas, for example, for the lowest level $S_{t_0}(0)=2.2120$. Straightforward but lengthy analytic calculations expand Eq.~\eqref{EntropyXlimits1_Small} up to the quadratic terms as:
\begin{equation}\tag{32a$'$}\label{eq:32a'}
S_{x_n}(\Lambda)=\ln2-1+2\Lambda-2\Lambda^2,\quad|\Lambda|\ll1,\quad n=0,1,\ldots,
\end{equation}
which is again a level-independent quantity. However, similar expansion, which, as our numerical results indicate, varies with the quantum number $n$, of Eq.~\eqref{EntropyKlimits1_Small} is hardly possible.

Momentum entropies of the two lowest split-off levels monotonically increase with the decrease of the absolute value of the negative Robin distance and at the very small $\Lambda$ their expressions are written as:
\begin{equation}\label{EntropyKlimits1_MinusZero}
S_{k_{e,o}}(\Lambda)=3\ln2+\ln\pi-1-\ln|\Lambda|,\quad\Lambda\rightarrow-0,
\end{equation}
what, in combination with Eq.~\eqref{EntropyXlimits1_MinusZero}, yields the following value of the total entropies:
\begin{equation}\label{EntropyTotallimits1_MinusZero}
S_{t_{e,o}}(\Lambda)=3\ln2+\ln\pi\approx3.2242,\quad\Lambda\rightarrow-0,
\end{equation}
satisfying, of course, inequality \eqref{EntropicInequality1}. As inset in panel (b) shows, the entropy $S_{t_0}$ of the lowest level approaches this $\Lambda$ independent limit in a nonmonotonic way passing at $\Lambda=-0.144$ through the maximum of $3.2305$.

\subsection{Onicescu energy and complexity CGL}\label{SubSec_Onicescu}
It is elementary to prove that the position Onicescu energy for any permutation of the Neumann and Dirichlet BCs [except the lowest Neumann state with $O_{x_0}(\infty)=1$] is equal to $3/2$. Eqs.~\eqref{OnicescuXlimits1_Small} and \eqref{OnicescuXlimits1_Large} show that tiny deviations from these limiting values are proportional to the first power of the small perturbation $\Lambda$ or $\Lambda^{-1}$ [or to its square for the near Neumann orbital with $n=0$, Eq.~\eqref{OnicescuXlimits1_Large}]. A comparison between either of these equations with their Shannon counterparts, Eqs.~\eqref{EntropyXlimits1_Small} and \eqref{EntropyXlimits1_Large}, respectively, reveals that in these two asymptotic regimes the two quantum information measures change in the opposite directions. This rule is further confirmed by panel (a) of Fig.~\ref{OnicescuSymmetricFig1}, which shows that for the ground (at $\Lambda=+\infty$) state the Onicescu energy has the only maximum of $1.8090$ at $\Lambda=-0.134$ (cf. with the corresponding data for the Shannon entropy from first paragraph of subsection ~\ref{SubSec_Shannon}) whereas for the higher-lying orbitals this $n$-dependent extremum is accompanied by the minimum at the positive Robin lengths. The magnitudes of the extreme values decrease with the quantum number and their location on the $\Lambda$ axis is  shifted closer to the Dirichlet BC for the greater $n$. Similar to the reasoning about the analogous behavior of the Shannon entropies, this is explained by the smaller influence of the BCs on the states with the larger energies.

Asymptotic expansions of the Shannon entropies and Onicescu energies near the Dirichlet and Neumann BCs yield expressions from Eqs.~\eqref{CGLXlimits1_Small} and \eqref{CGLXlimits1_Large} for the position complexity $CGL_x$ in these two limiting cases, where $3/e\approx1.1036$.

\begin{figure}
\centering
\includegraphics[width=\columnwidth]{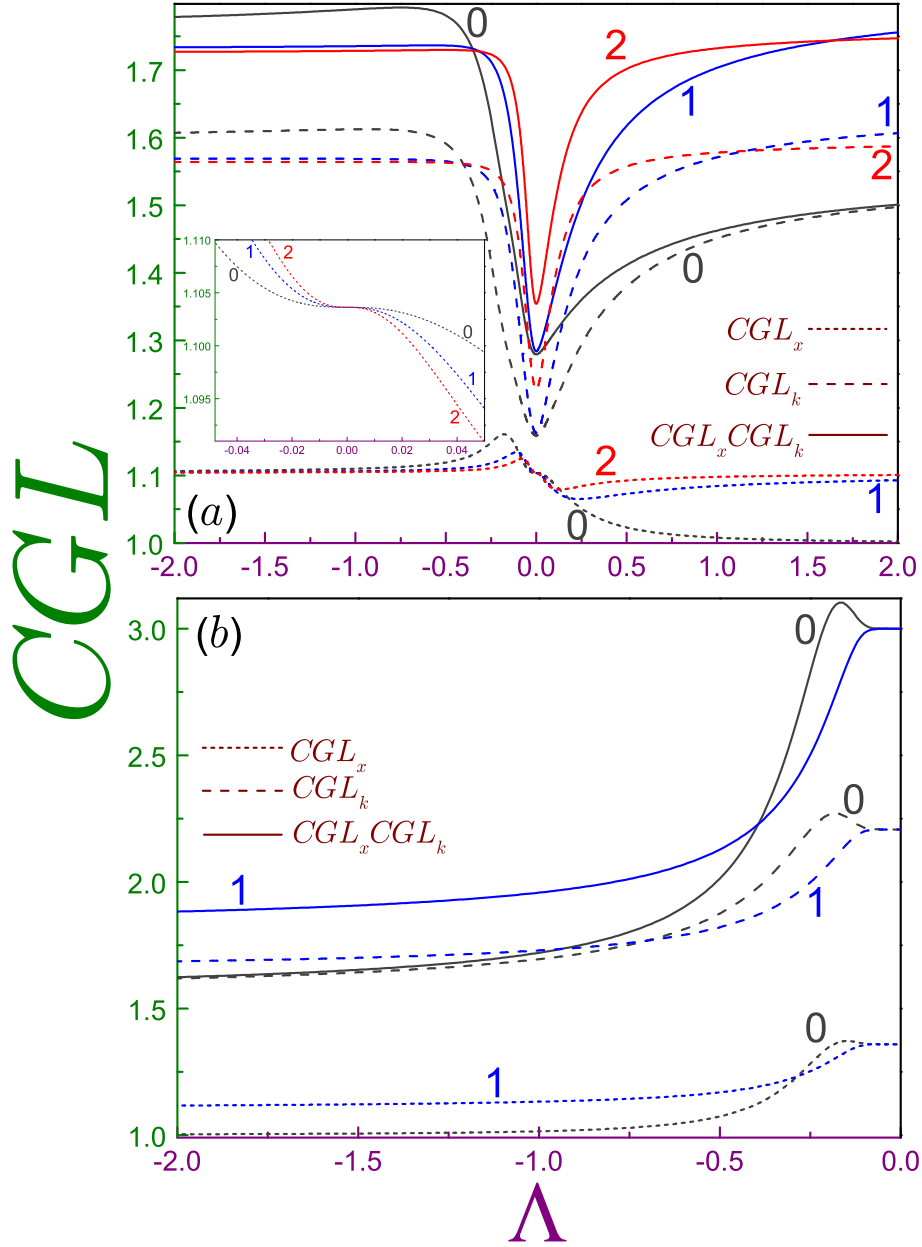}
\caption{\label{CGLsymmetricFig1}
Statistical complexities $CGL$ of the symmetric QW as functions of the Robin distance $\Lambda$. The same convention as in Figs.~\ref{EntropySymmetricFig1} and~\ref{OnicescuSymmetricFig1} is used. Inset in panel (a) shows an enlarged view of the formation of the position plateau around the Dirichlet BC.}
\end{figure}

$CGL_x$ variation with the Robin distance is depicted by the dotted lines in Fig.~\ref{CGLsymmetricFig1}. Its most remarkable feature are plateaus around the Dirichlet BC. Enlarged view of these $\Lambda$-independent parts of the complexity presented in the inset of panel (a) shows that their widths decrease at the higher quantum numbers. A formation of the flat region can be seen from the expressions for the corresponding Shannon entropy, Eq.~\eqref{EntropyXlimits1_Small}, and Onicescu energy, Eq.~\eqref{OnicescuXlimits1_Small}, which change in the opposite directions and, when amalgamated into the complexity $CGL_x$, produce the $\Lambda$-independent Dirichlet value $3/e$. Moreover, similar to the Shannon entropy, expanding Eq.~\eqref{OnicescuXlimits1_Small} up to the quadratic terms:
\begin{equation}\tag{34a$'$}\label{eq:34a'}
O_{x_n}(\Lambda)=\frac{3}{2}-3\Lambda+6\Lambda^2,\quad|\Lambda|\ll1,\quad n=0,1,\ldots,
\end{equation}
and combining it with Eq.~\eqref{eq:32a'}, one sees that in this higher-order approximation the product $e^{S_{x_n}(\Lambda)}O_{x_n}(\Lambda)$ stays constant too in the vicinity of the Dirichlet BC. Physical reason of this phenomenon lies in the fact that the variation of one quantum information measure comprising the complexity is exactly compensated by the reverse alteration of the second constituent. 

Position Onicescu energies of the two split-off orbitals monotonically increase with the de Gennes distance varying from the almost Neumann BC, Eq.~\eqref{OnicescuXlimits1_Large}, to the negative zero, Eq.~\eqref{OnicescuXlimits1_MinusZero}, when they unrestrictedly increase, as panel (b) of Fig.~\ref{OnicescuSymmetricFig1} demonstrates. Accordingly, the complexities $CGL$ tend in the latter limit to $e/2\approx1.3591$, see panel (b) of Fig.~\ref{CGLsymmetricFig1}. It is instructive to underline that, as expected, all position complexities as well as their momentum counterparts discussed below do obey inequality~\eqref{CGLinequality1}, which comes to its saturation for the lowest Neumann level according to the dependence from Eq.~\eqref{CGLXlimits1_Large}.

Turning to the discussion of the momentum disequilibria $O_{k_n}(\Lambda)$, we first provide analytic expressions for the Dirichlet and Neumann Onicescu energies:
\begin{subequations}\label{OnicescuK_DirNeu1}
\begin{eqnarray}\label{OnicescuK_Dir1}
O_{k_n}(0)&=&\frac{15+2\pi^2(n+1)^2}{12\pi^3(n+1)^2},\quad n=0,1,2,\ldots\\
\label{OnicescuK_Neu1}
O_{k_n}(\infty)&=&\left\{\begin{array}{cc}
\frac{1}{3\pi},&n=0\\
\frac{3+2\pi^2n^2}{12\pi^3n^2},&n\geq1.
\end{array}
\right.
\end{eqnarray}
\end{subequations}
For high-lying states, they saturate from above to the level-independent constant:
\begin{equation}\label{OnicescuK_DirNeu_LimitLargeN1}
O_{k_n}(0)=O_{k_n}(\infty)=\frac{1}{6\pi},\quad n\rightarrow\infty.
\end{equation}
It follows from Eqs.~\eqref{OnicescuK_DirNeu1} that the Dirichlet and Neumann momentum disequilibria are decreasing functions of the quantum index $n$. The same holds true for their product with their position counterpart $O_{x_n}O_{k_n}$:
\begin{subequations}\label{OnicescuXK_DirNeu1}
\begin{eqnarray}\label{OnicescuXK_Dir1}
O_{x_n}(0)\,O_{k_n}(0)&=&\frac{15+2\pi^2(n+1)^2}{8\pi^3(n+1)^2},\quad n=0,1,2,\ldots\\
\label{OnicescuXK_Neu1}
O_{x_n}(\infty)\,O_{k_n}(\infty)&=&\left\{\begin{array}{cc}
\frac{1}{3\pi},&n=0\\
\frac{3+2\pi^2n^2}{8\pi^3n^2},&n\geq1.
\end{array}
\right.
\end{eqnarray}
\end{subequations}
Eqs.~\eqref{OnicescuXK_DirNeu1} show that $n$th Dirichlet product $O_xO_k$ is greater than its $n$ and all the more $(n+2)$th Neumann counterparts. Accordingly, one can expect at least its one extremum as the Robin distance changes from the large negative to positive lengths. Utilizing Eq.~\eqref{FunctionsSymmetricLimit1_Small} for finding corrections to the Dirichlet momentum waveform disturbed by small $\Lambda$ perturbation, one derives:
\begin{equation}\label{OnicescuKlimits1_Small}
O_{k_n}(\Lambda)=O_{k_n}(0)(1+2\Lambda),\quad|\Lambda|\ll1,
\end{equation}
leading to the cancellation of the linear term in the corresponding product. The fact that the minimum power of the admixture to $O_{x_n}(0)O_{k_n}(0)$ is quadratic means that the Dirichlet BC is the extremum point of this dependence. As solid lines in Fig.~\ref{OnicescuSymmetricFig1} manifest, it is a global maximum supplemented by the broad minimum whose location on the negative $\Lambda$ semi axis is shifted closer to zero with the increase of the quantum index. This lowest value of the product is a consequence of the corresponding extremum of the momentum Onicescu energy whereas its $n$-dependent maximum lies at the positive extrapolation lengths.

Such $O_k$ behavior has its drastic consequences on the momentum complexity $CGL_{k_n}(\Lambda)$; namely, a comparison of Eqs.~\eqref{EntropyKlimits1_Small} and \eqref{OnicescuKlimits1_Small} immediately reveals that it has a minimum at the Dirichlet BC, and, since a position measure $CGL_{x_n}(\Lambda)$ is flat at $|\Lambda|\ll1$, as discussed above, this extremum at $\Lambda=0$ is also characteristic for the product of the two complexities $CGL_{x_n}(\Lambda)CGL_{k_n}(\Lambda)$. Fig.~\ref{CGLsymmetricFig1} demonstrates that the minimal values $CGL_{k_n}(0)$ and $CGL_{x_n}(0)CGL_{k_n}(0)$ increase with $n$.

Momentum Onicescu energies of the two split-off levels monotonically decrease when the Robin distance approaches zero from the left, and at the very small extrapolation lengths they fade as:
\begin{equation}\label{OnicescuKlimits1_MinusZero}
O_{k_{e,o}}(\Lambda)=\frac{3}{4\pi}|\Lambda|,\quad\Lambda\rightarrow-0,
\end{equation}
what, in combination with Eq.~\eqref{OnicescuXlimits1_MinusZero}, yields the $\Lambda$-independent product of the position and momentum disequilibria:
\begin{equation}\label{OnicescuXKlimits1_MinusZero}
O_{x_{e,o}}(\Lambda)O_{k_{e,o}}(\Lambda)=\frac{3}{8\pi}\approx0.1194,\quad\Lambda\rightarrow-0.
\end{equation}
This, in turn, leads to the  Robin-length-independent expressions for the complexities $CGL$ at $\Lambda\rightarrow-0$:
\begin{subequations}\label{CGL1_MinusZero}
\begin{eqnarray}\label{CGL1K_MinusZero}
CGL_{k_{e,o}}(\Lambda)&=&6/e\approx2.2073\\
\label{CGL1XK_MinusZero}
CGL_{x_{e,o}}(\Lambda)CGL_{k_{e,o}}(\Lambda)&=&3.
\end{eqnarray}
\end{subequations}
Similar to the sum of the Shannon entropies, the lowest level position and momentum measures and their product approach these limits nonmonotonically: the corresponding maxima of $1.3715$, $2.2712$ and $3.1040$ are achieved at $\Lambda=-0.150$, $-0.182$ and $-0.166$, respectively.

\subsection{Fisher information}\label{SubSec_Fisher}
Fig.~\ref{FisherSymmetricFig1}(a) shows that position Fisher information of the extended states basically repeats energy dependence on the extrapolation length, as it follows from Eq.~\eqref{FisherX_Anal1}. Its momentum counterpart while changing its value from $I_{k_{n+2}}(-\infty)$ to $I_{k_n}(\infty)$ passes through the minimum whose location on the negative semi axis moves closer to $\Lambda=0$ at the increasing index $n$, see panel (b) of Fig.~\ref{FisherSymmetricFig1}. Smaller values of the Fisher information mean, according to Eq.~\eqref{Fisher1}, less oscillatory structure of the associated functions. The product of the position and momentum Fisher informations generally repeats the shape of the former one, as solid lines in panel (a) demonstrate. As our numerical results reveal, for the ground state Eq.~\eqref{FisherUncertainty1} does not hold true at $\Lambda>0.062$.

\begin{figure}
\centering
\includegraphics[width=\columnwidth]{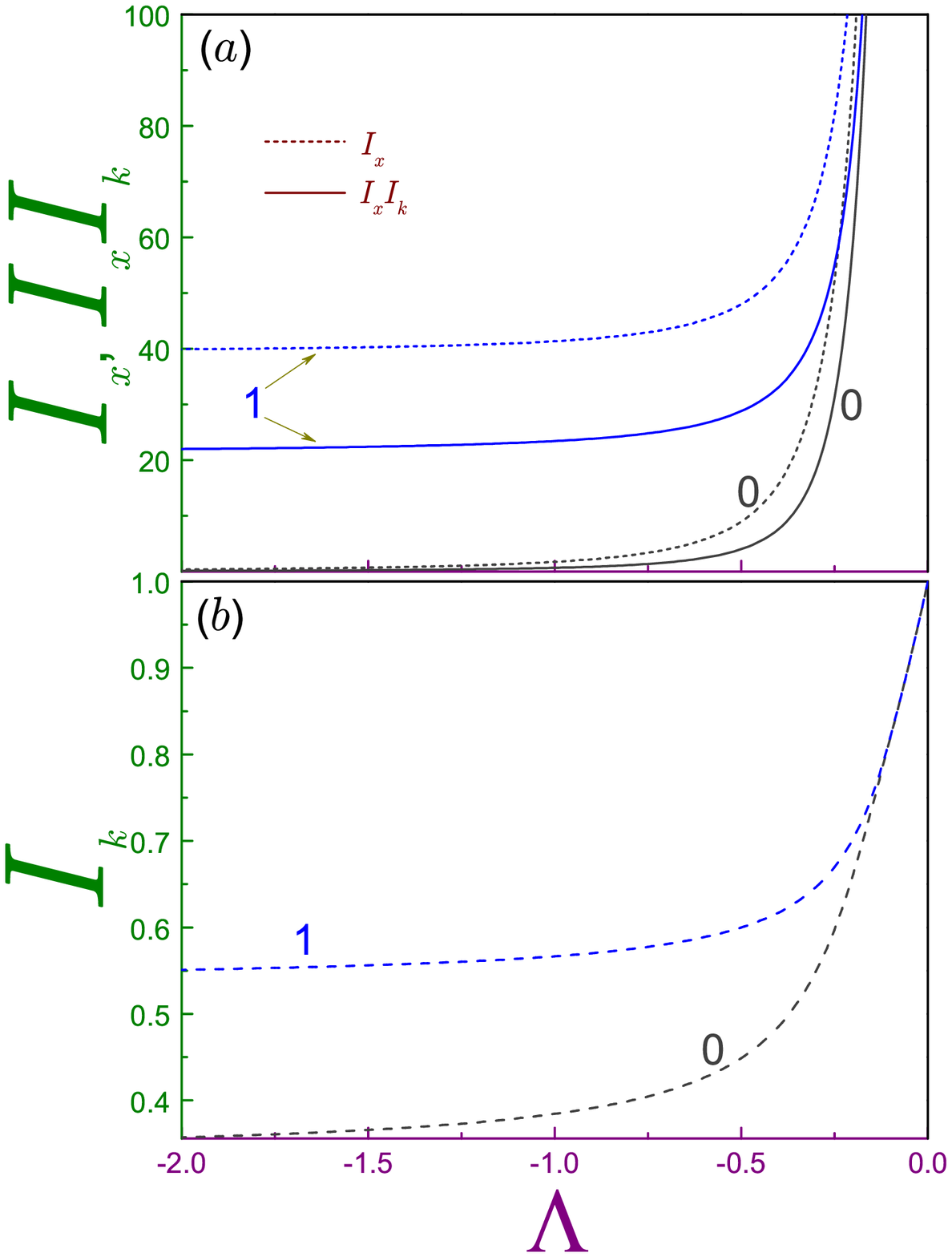}
\caption{\label{FisherSymmetricFig2}
The same as in Fig.~\ref{FisherSymmetricFig1} but for the two split-off states.}
\end{figure}

An increasing localization of the two split-off levels at the surfaces $x=\pm1/2$  with the negative Robin length approaching zero means greater steepness of the position functions $\Psi(x)$, see panels (a) and (b) of Fig.~\ref{FunctionsXsymmetricFig1}. This, in turn, results in the growing Fisher information, as Fig.~\ref{FisherSymmetricFig2}(a) demonstrates. At the very small negative extrapolation distances, they unrestrictedly rise, according to Eq.~\eqref{FisherXlimits1_MinusZero}. At the same time, even though momentum functions fade to zero, according to Eq.~\eqref{MomentumDensityAtZero1} and panels (a) and (b) of Fig.~\ref{DensityKsymmetricFig1}, their oscillatory structure still persists, as it follows from Eq.~\eqref{MomentumDensityAtZero1}. As a result, their momentum Fisher informations do not disappear with the negative extrapolation length approaching zero but tend to unity:
\begin{equation}\label{FisherK5}
I_{k_{e,o}}(\Lambda)=1-2|\Lambda|+2|\Lambda|^2,\quad\Lambda\rightarrow-0,
\end{equation}
what can be easily shown from Eq.~\eqref{FisherK3} adapted for $E<0$. This means that in this regime the product of the position and momentum informations is growing as an inverse of the square of the Robin distance:
\begin{equation}\label{FisherK6}
I_{x_{e,0}}(\Lambda)I_{k_{e,0}}(\Lambda)=\frac{4}{|\Lambda|^2},\quad\Lambda\rightarrow-0.
\end{equation}

\section{Asymmetric well: $\Lambda_-=-\Lambda_+=\Lambda$}\label{Sec_Asymmetric}
A characteristic feature of this geometry is the independence of the energies of the excited states on the extrapolation length that can not affect the Dirichlet spectrum. Even more spectacular is the fact that, despite the variation of the waveforms with the de Gennes parameter, the associated position quantum-information measures are not altered either:
\begin{eqnarray}\label{ShannonXAsym1}
S_{x_n}(\Lambda)&=&\ln2-1\\
\label{FisherXAsym1}
I_{x_n}(\Lambda)&=&4\pi^2n^2\\
\label{OnicescuXAsym1}
O_{x_n}(\Lambda)&=&\frac{3}{2}\\
\label{CGLXAsym1}
CGL_{x_n}(\Lambda)&=&\frac{3}{e}
\end{eqnarray}
and are equal to their Dirichlet values too. Thus, for the excited states the only result of varying of the Robin distance $\Lambda$ is the deformation of the wave functions that, nevertheless, is unable to change neither energies nor any of the position measures discussed here. For the ground level, however, a situation is completely different; for example, its position Shannon entropy reads:
\begin{subequations}\label{ShannonX0Asym1}
\begin{align}\label{ShannonX0Asym1_General}
S_{x_0}(\Lambda)&=\ln\!\left(\!\Lambda\sinh\frac{1}{\Lambda}\right)-\frac{1}{\Lambda}\coth\frac{1}{\Lambda}+1\\
\intertext{with the asymptotes:}
\label{ShannonX0Asym1_LargeLamda}
S_{x_0}(\Lambda)&=-\frac{1}{6\Lambda^2}+\frac{1}{60\Lambda^4}+\ldots,\quad\Lambda\rightarrow\infty\\
\label{ShannonX0Asym1_SmallLamda}
S_{x_0}(\Lambda)&=\ln\Lambda+1-\ln2,\quad\Lambda\rightarrow0.
\end{align}
\end{subequations}
Note that the coefficient of the quadratic admixture of $1/\Lambda$ to the Neumann Shannon entropy is different from that for the symmetric BC, Eq.~\eqref{EntropyXlimits1_Large}, whereas in the opposite limit Eq.~\eqref{ShannonX0Asym1_SmallLamda} does coincide with the expression for the only bound level of the single attractive Robin wall~\cite{Olendski3}. Ground-state Fisher information is at any $\Lambda$ exactly the same as its counterpart for the latter geometry \cite{Olendski3}:
\begin{equation}\label{FisherX0Asym1}
I_{x_0}(\Lambda)=\frac{4}{\Lambda^2},
\end{equation}
while the Onicescu energy is:
\begin{subequations}\label{OnicescuX0Asym1}
\begin{align}\label{OnicescuX0Asym1_General}
O_{x_0}(\Lambda)&=\frac{1}{\Lambda}\coth\frac{1}{\Lambda},\\
\intertext{and its limits are:}
\label{OnicescuX0Asym1_LargeLamda}
O_{x_0}(\Lambda)&=1+\frac{1}{3\Lambda^2}-\frac{1}{45\Lambda^4}+\ldots,\quad\Lambda\rightarrow\infty,\\
\label{OnicescuX0Asym1_SmallLamda}
O_{x_0}(\Lambda)&=\frac{1}{\Lambda},\quad\Lambda\rightarrow0.
\end{align}
\end{subequations}
Last equation shows that the lowest-level position disequilibrium is equal to that of the single surface \cite{Olendski3} in the limit of the vanishing lengths $\Lambda$ only. From the above formulae, one can easily derive the expression for the shape complexity $CGL_{x_0}$ but we provide here its asymptotics only:
\begin{subequations}\label{CGLX0Asym1}
\begin{eqnarray}
\label{CGLX0Asym1_LargeLamda}
CGL_{x_0}(\Lambda)&=&1+\frac{1}{6\Lambda^2},\quad\Lambda\rightarrow\infty,\\
\label{CGLX0Asym1_SmallLamda}
CGL_{x_0}(\Lambda)&=&\frac{e}{2},\quad\Lambda\rightarrow0.
\end{eqnarray}
\end{subequations}
It is instructive to point out that, similar to the same Robin distances at the both walls, Sec.~\ref{Sec_Symmetric1}, the complexity approaches its Neumann saturation limit of inequality~\eqref{CGLinequality1} as the inverse square of the large extrapolation length, Eq.~\eqref{CGLX0Asym1_LargeLamda}, but the coefficient at $\Lambda^{-2}$ is different from the symmetric geometry, Eq.~\eqref{CGLXlimits1_Large}.

Ground-state momentum function $\Phi_0(\Lambda;k)$ is written explicitly as
\begin{subequations}\label{FunctionK0Asym1}
\begin{align}
\Phi_0(\Lambda;k)&=\left(\frac{2}{\pi}\frac{\Lambda}{\sinh\frac{1}{\Lambda}}\right)^{1/2}\nonumber\\
\label{FunctionK0Asym1_1}
&\times\frac{\sinh\frac{1}{2\Lambda}\cos\frac{k}{2}-i\cosh\frac{1}{2\Lambda}\sin\frac{k}{2}}{1-i\Lambda k},\\
\intertext{and its limiting cases are:}
\Phi_0(\Lambda;k)=&\left(\frac{2}{\pi}\right)^{1/2}\frac{\sin\frac{k}{2}}{k}\nonumber\\
\label{FunctionK0Asym1_Large}
+&\frac{i}{\Lambda k}\left(\frac{2}{\pi}\right)^{1/2}\left(\frac{1}{2}\cos\frac{k}{2}-\frac{\sin\frac{k}{2}}{k}\right),\quad\Lambda\rightarrow\infty,\\
\label{FunctionK0Asym1_Small}
\Phi_0(\Lambda;k)=&\left(\frac{\Lambda}{\pi}\right)^{1/2}\frac{e^{-ik/2}}{1-i\Lambda k},\,\Lambda\rightarrow0.
\end{align}
\end{subequations}
The first term in the right-hand side of Eq.~\eqref{FunctionK0Asym1_Large} is just the Neumann QW lowest-level momentum waveform \cite{Olendski2,Bialynicki2} and second item there is a small admixture due to finiteness of the extrapolation length whereas the opposite asymtpote differs from the corresponding expression for the single attractive Robin wall \cite{Olendski3} by the complex phase factor $e^{-ik/2}$ only what means that in this regime the particle strongly localized near one surface does not 'feel' the influence of the second interface, and all its characteristics will be similar to those discussed earlier \cite{Olendski3}:
\begin{subequations}\label{ShannonK0Asym1}
\begin{eqnarray}\label{ShannonK0Asym1_SmallLamda}
S_{k_0}(\Lambda)&=&2\ln2+\ln\pi-\ln\Lambda\\
\label{ShannonT0Asym1_SmallLamda}
S_{x_0}(\Lambda)+S_{k_0}(\Lambda)&=&1+\ln\pi+\ln2,
\end{eqnarray}
\end{subequations}
\begin{subequations}\label{FisherK0Asym1}
\begin{eqnarray}\label{FisherK0Asym1_SmallLamda}
I_{k_0}(\Lambda)&=&\frac{\Lambda^2}{2}\\
\label{FisherT0Asym1_SmallLamda}
I_{x_0}(\Lambda)I_{k_0}(\Lambda)&=&2,
\end{eqnarray}
\end{subequations}
\begin{subequations}\label{OnicescuK0Asym1}
\begin{eqnarray}\label{OnicescuK0Asym1_SmallLamda}
O_{k_0}(\Lambda)&=&\frac{\Lambda}{2\pi}\\
\label{OnicescuT0Asym1_SmallLamda}
O_{x_0}(\Lambda)O_{k_0}(\Lambda)&=&\frac{1}{2\pi},
\end{eqnarray}
\end{subequations}
\begin{subequations}\label{CGLK0Asym1}
\begin{eqnarray}\label{CGLK0Asym1_SmallLamda}
CGL_{k_0}(\Lambda)&=&2\\
\label{CGLT0Asym1_SmallLamda}
CGL_{x_0}(\Lambda)CGL_{k_0}(\Lambda)&=&e,
\end{eqnarray}
\end{subequations}
$\Lambda\rightarrow0$. In the opposite limit of $1/\Lambda\ll1$ one gets:
\begin{subequations}
\begin{eqnarray}
O_{k_0}(\Lambda)=\frac{1}{3\pi}-\frac{2}{45\pi}\frac{1}{\Lambda^2}\\
O_{x_0}(\Lambda)O_{k_0}(\Lambda)=\frac{1}{3\pi}+\frac{1}{15\pi}\frac{1}{\Lambda^2}.
\end{eqnarray}
\end{subequations}
Note that the expression for $O_{k_0}(\Lambda)$ at the arbitrary $\Lambda$ can be derived analytically but since it contains many terms we do not write it here.

\begin{figure}
\centering
\includegraphics[width=0.8\columnwidth]{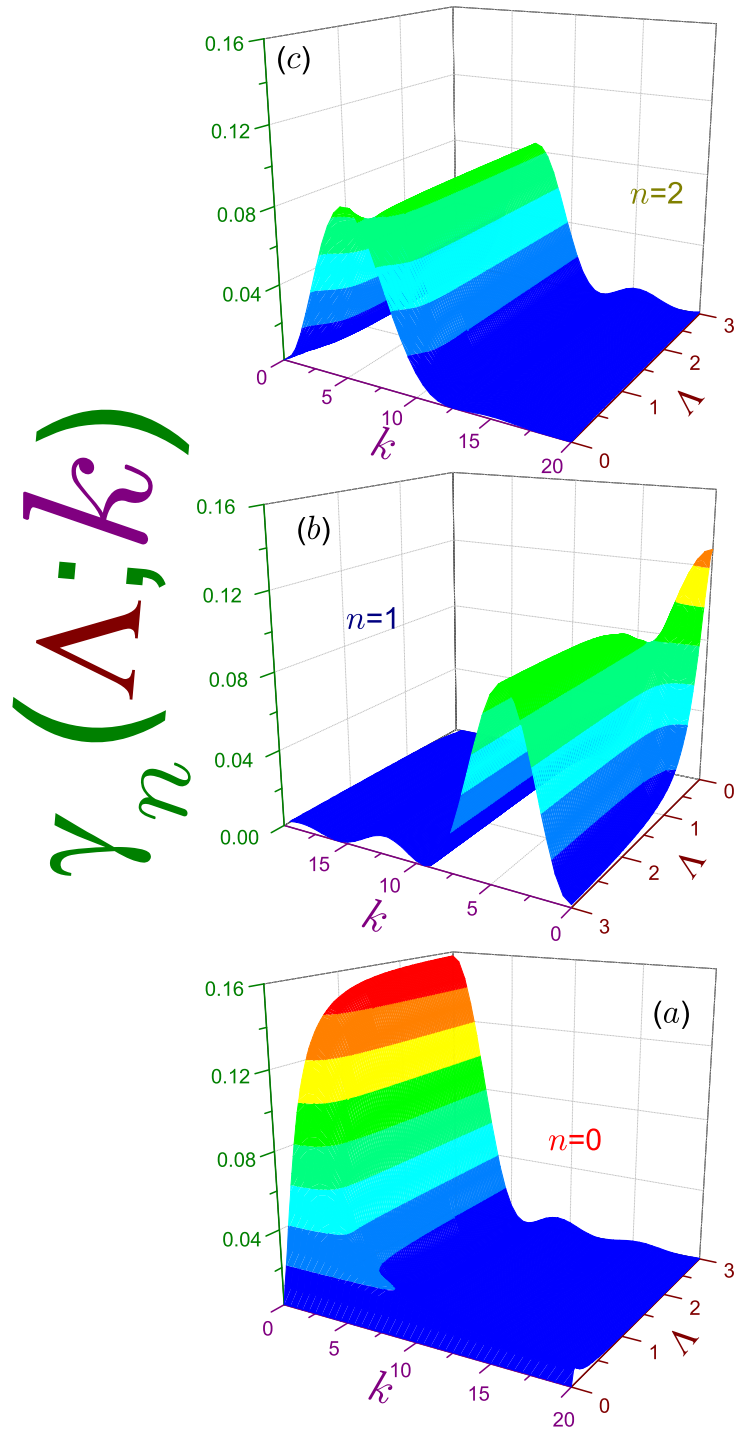}
\caption{\label{DensityKasymmetricFig1}
Momentum wave functions $\gamma_n(\Lambda;k)$ of the asymmetric QW in terms of the momentum $k$ and extrapolation length $\Lambda$ for (a) the lowest level, $n=0$, (b) first excited orbital, $n=1$, and (c) second excited state, $n=2$. Different viewing perspective is used in panel (b) as compared to subplots (a) and (c).}
\end{figure}

Part (a) of Fig.~\ref{DensityKasymmetricFig1} showing a 3D plot of the ground-state momentum density
\begin{equation}\label{DensityK0Asym1}
\gamma_0(\Lambda;k)=\frac{2}{\pi}\frac{\Lambda}{\sinh\frac{1}{\Lambda}}\frac{\cosh^2\frac{1}{2\Lambda}-\cos^2\frac{k}{2}}{1+\Lambda^2k^2},
\end{equation}
demonstrates that its maximum monotonically decreases with the extrapolation length and, at the very small Robin distances, the oscillations are completely subdued with the function turning to zero together with $\Lambda$:
\begin{equation}\label{DensityK0Asym2}
\gamma_0(\Lambda;k)=\frac{\Lambda}{\pi}\frac{1}{1+\Lambda^2k^2},\quad\Lambda\rightarrow0,
\end{equation}
as it follows from Eq.~\eqref{FunctionK0Asym1_Small}. Two upper panels of the figure exhibit transformations from the Neumann to Dirichlet momentum densities for the two lowest excited levels as de Gennes distance varies. Corresponding waveforms are:
\begin{subequations}\label{FunctionsMomentumAsymmetric1}
\begin{eqnarray}
\Phi_{2n}(\Lambda;k)&=&(-1)^n\left[\frac{\pi}{1+(2\pi n\Lambda)^2}\right]^{1/2}\nonumber\\
\label{FunctionsMomentumAsymmetric1_Even}
&\times&\frac{4n(i+\Lambda k)}{k^2-(2\pi n)^2}\sin\frac{k}{2},\,n=1,2,\ldots\\
\Phi_{2n+1}(\Lambda;k)&=&(-1)^n\left(\frac{\pi}{1+[\pi(2n+1)\Lambda]^2}\right)^{1/2}\nonumber\\
\label{FunctionsMomentumAsymmetric1_Odd}
&\times&\frac{2(2n+1)(i+\Lambda k)}{k^2-[\pi(2n+1)]^2}\cos\frac{k}{2},\,n=0,1,\ldots.
\end{eqnarray}
\end{subequations}
\begin{figure}
\centering
\includegraphics[width=\columnwidth]{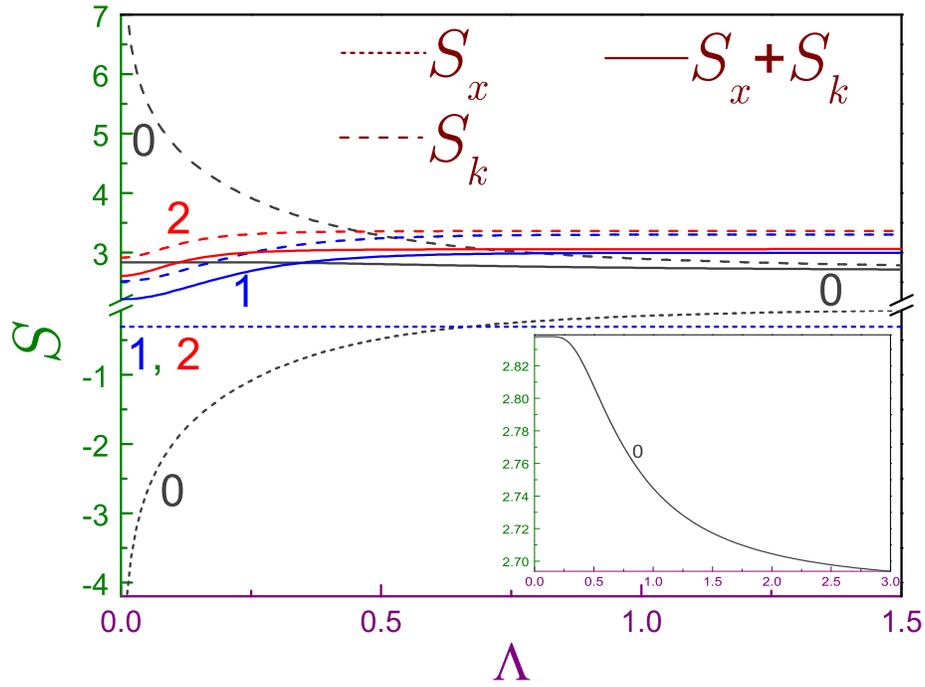}
\caption{\label{EntropyAsymmetricFig1}
Position $S_{x_n}$ (dotted lines) and momentum $S_{k_n}$ (dashed curves) Shannon entropies together with their sum $S_{x_n}+S_{k_n}$ (solid lines) of the asymmetric QW as functions of the extrapolation length $\Lambda$. Numbers near the curves denote corresponding quantum indexes $n$. Note vertical line break from $-0.07$ to $2.2$. Inset shows an enlarged view of $S_{t_0}(\Lambda)$ for the ground level.}
\end{figure}

Shannon entropies of the three lowest levels are shown in Fig.~\ref{EntropyAsymmetricFig1}. At the infinitely small extrapolation lengths the position and momentum components of the ground state diverge as positive and negative natural logarithm of the Robin distance, see Eqs.~\eqref{ShannonX0Asym1_SmallLamda} and~\eqref{ShannonK0Asym1_SmallLamda}, in such a way that their sum stays finite and equal to $2.8379$, as it follows from Eq.~\eqref{ShannonT0Asym1_SmallLamda}. The increase of the de Gennes parameter decreases $S_{t_0}(\Lambda)$, which at the Neumann limit, $\Lambda\rightarrow\infty$, consists mainly from the momentum component with negligible contribution from $S_{x_0}(\Lambda)$, Eq.~\eqref{ShannonX0Asym1_LargeLamda}, and approaches $2.6834$ \cite{Olendski2}. Note that at $0\leq\Lambda\lesssim0.3$, the ground-state sum almost does not depend on the Robin length forming a plateau, as inset shows. Formation of this flat plat at quite small extrapolation distances is a reflection of the fact that at strong enough confinement a further increase of the binding can not change the value of $S_t$ since each of its items is already in the asymptotic regime, Eqs.~\eqref{ShannonX0Asym1_SmallLamda} and~\eqref{ShannonK0Asym1_SmallLamda}. For any excited state, $n\geq1$, the change in $S_{x_n}(\Lambda)+S_{t_n}(\Lambda)$ is due to the corresponding variation of its momentum part only since the position entropy stays constant. It was established before \cite{Olendski2} that at the fixed principal quantum number the Dirichlet total entropy is smaller than its Neumann counterpart and since both of them are increasing functions of $n$, the sum $S_{t_n}(\Lambda)$ grows with the extrapolation length, as Fig.~\ref{EntropyAsymmetricFig1} demonstrates.

\begin{figure}
\centering
\includegraphics[width=\columnwidth]{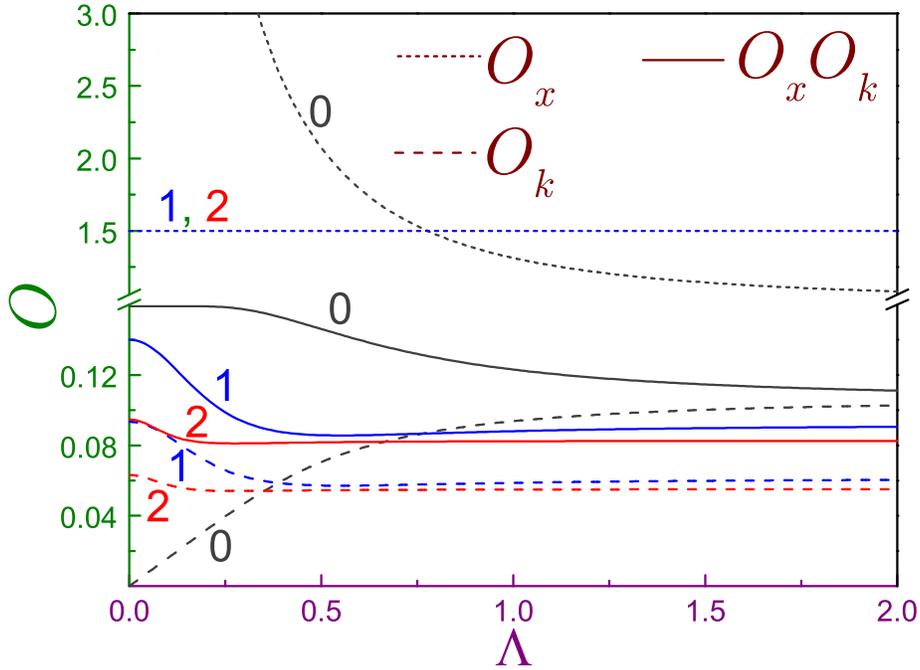}
\caption{\label{OnicescuAsymmetricFig1}
Position $O_{x_n}$ (dotted lines) and momentum $O_{k_n}$ (dashed curves) disequilibria together with their product $O_{x_n}O_{k_n}$ (solid lines) of the asymmetric QW as functions of the extrapolation length $\Lambda$. Numbers near the curves denote corresponding quantum indexes $n$. There is a vertical line break from $0.16$ to $1.06$.}
\end{figure}

\begin{figure}
\centering
\includegraphics[width=\columnwidth]{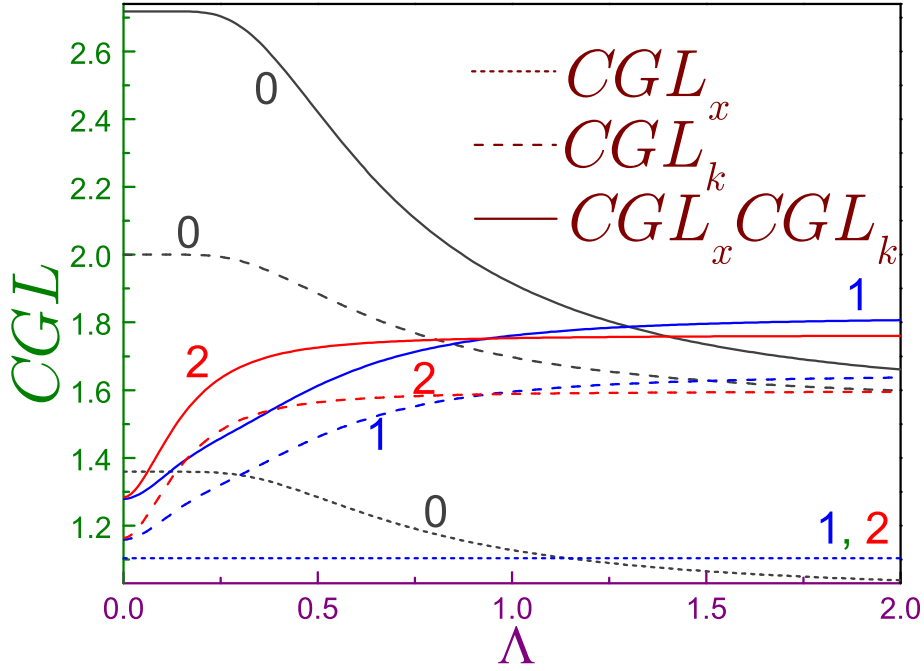}
\caption{\label{CGLAsymmetricFig1}
Position $CGL_{x_n}$ (dotted lines) and momentum $CGL_{k_n}$ (dashed curves) complexities $CGL$ together with their product $CGL_{x_n}CGL_{k_n}$ (solid lines) of the asymmetric QW as functions of the extrapolation length $\Lambda$. Numbers near the curves denote corresponding quantum indexes $n$.}
\end{figure}

Plateau at small Robin distances is also characteristic for the product of the ground-state Onicescu energies $O_{x_0}(\Lambda)O_{k_0}(\Lambda)$, see Fig.~\ref{OnicescuAsymmetricFig1}, and, as a result, to the complexities $CGL_{x_0}(\Lambda)CGL_{k_0}(\Lambda)$, Fig.~\ref{CGLAsymmetricFig1}, with the corresponding values of $1/(2\pi)$ and $e$. It has to be noted that both position and momentum products of $e^SO$ of the lowest level do not change either for the variation of the quite small extrapolation lengths staying equal to $e/2$ and $2$, respectively. For the arbitrary quantum index, the product of two disequilibria $O_{x_n}(\Lambda)O_{k_n}(\Lambda)$ is a monotonically decreasing function of the Robin distance while for the complexity $CGL$ this is true for the ground state only. And, of course, both position and momentum complexities for this BC geometry satisfy inequality \eqref{CGLinequality1} too.

Momentum Fisher informations of the excited states can be calculated analytically but the expressions are too bulky and not written here. Their asymptotes are:
\begin{equation}\label{FisherK8}
I_{k_n}(\Lambda)=\frac{1}{3}\left(1\mp\frac{6}{\pi^2n^2}\right)\pm\left\{
\begin{array}{cc}
4\pi^2n^2\Lambda^4,&\Lambda\ll1,\\
\frac{16}{\pi^4n^4}\frac{1}{\Lambda^2},&\frac{1}{\Lambda}\ll1
\end{array}
\right\},\,n=1,2,\ldots.
\end{equation}
Note that due to the different BC geometries, tiny admixtures for the antisymmetric configuration are different from those provided in Eqs.~\eqref{FisherK4} for the equal extrapolation lengths at both interfaces; namely, the lowest nonzero power of the Dirichlet (Neumann) perturbation is quartic (quadratic) in the former case while for the latter arrangement they depend linearly on the small disturbance.

\begin{figure}
\centering
\includegraphics[width=\columnwidth]{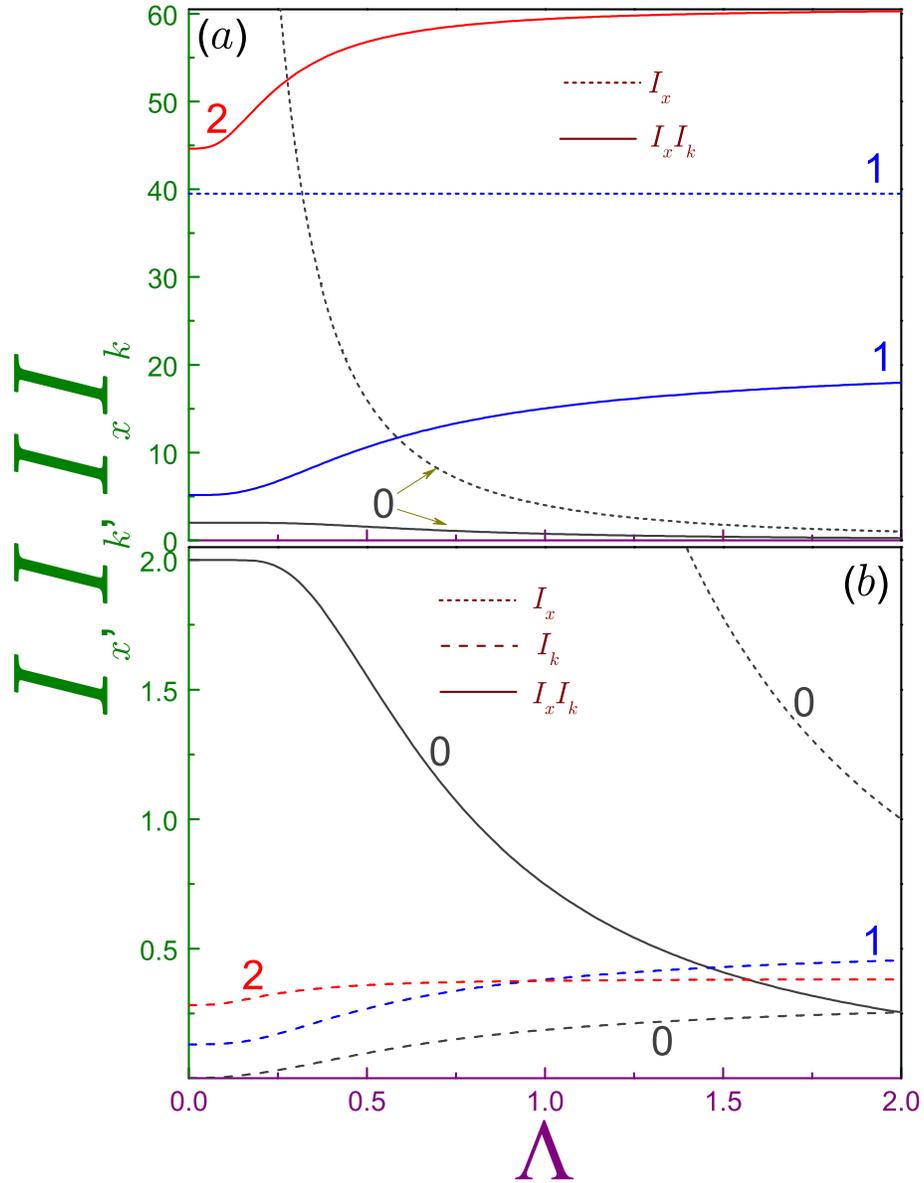}
\caption{\label{FisherAsymmetricFig1}
Position $I_{x_n}$ (dotted lines) and momentum $I_{k_n}$ (dashed curves) Fisher informations together with their product $I_{x_n}I_{k_n}$ (solid lines) of the asymmetric QW as functions of the extrapolation length $\Lambda$. Numbers near the curves denote corresponding quantum indexes $n$. Note different vertical scales in panels (a) and (b).}
\end{figure}

Fig.~\ref{FisherAsymmetricFig1} shows evolution of the Fisher informations of the asymmetric QW. Similar to other quantum-information measures, a characteristic feature of the ground-state product $I_{x_0}(\Lambda)I_{k_0}(\Lambda)$ is a plateau at the small Robin distances with its value equal to $2$, Eq.~\eqref{FisherT0Asym1_SmallLamda} , whereas each of the multipliers changes in this regime according to Eqs.~\eqref{FisherX0Asym1} and \eqref{FisherK0Asym1_SmallLamda}. Observe that the vanishing momentum information at infinitely small lengths, Eq.~\eqref{FisherK0Asym1_SmallLamda}, has no resemblance whatever to its counterparts for the symmetric geometry, Eq.~\eqref{FisherK5} and Fig.~\ref{FisherSymmetricFig2}(b), what is explained by the different associated waveforms, see Eqs.~\eqref{DensityK0Asym2} and \eqref{MomentumDensityAtZero1}, respectively. It has to be noted also that for this level Eq.~\eqref{FisherUncertainty1} is violated at any Robin distances. Higher lying quantum states show a continuous transition from the Dirichlet BC at $\Lambda=0$ to the Neumann configuration at the infinitely large de Gennes parameter.

\section{Conclusions}\label{Sec_Conclusions}
Three major quantum-information measures - Shannon entropy, Fisher information and Onicescu energy - of the Robin QW have been analyzed for two geometries of the BCs. These functionals characterize different facets of the density distribution in the nanostructure and, thus, knowledge of one of them complements the data obtained from the analysis of the other measures. For the symmetric configuration of the BCs characterized by the extrapolation length $\Lambda$ it was shown that the Dirichlet requirement presents a special case; for example, the sum (product) of the position and momentum Shannon entropies (Onicescu energies) reaches at $\Lambda=0$ the minimum (maximum)  for the levels that exist at any sign of the Robin parameter. Since the quantum entropy obeys fundamental inequality, Eq.~\eqref{EntropicInequality1}, it means that the Dirichlet BC bears the most of the available information on the overall particle behavior or, in other words, from point of view of information theory it is closest to the classical mechanics. For the quantum states that exist at the negative extrapolation lengths only and whose energies diverge in the limiting case $\Lambda\rightarrow-0$ according to Eq.~\eqref{EnergySymmetricLimit1_MinusZero}, analytic expressions for the same asymptote have been derived for all measures showing, e.g., that position (momentum) Shannon entropy diverges as positive (negative) logarithm of the absolute value of the de Gennes distance and the sum satisfies Eq.~\eqref{EntropicInequality1}. A remarkable property of the asymmetric geometry with $\Lambda_-=-\Lambda_+\equiv\Lambda$ is the fact that for any excited orbital not only its energy stays constant as the Robin length varies along the whole axis but the same holds true for position component of any of the quantum-information measures. This curious $\Lambda$-independence takes place despite the deformation of the associated waveforms when the extrapolation parameter changes. Some other numerical and analytic results have been obtained and discussed and similarities and differences of the two BC configurations were pointed out.

Additionally, a complex measure $e^SO$ has been calculated too. This product jointly describes the uniformity and delocalization  facets of the particle probability density \cite{Esquivel1}. As a result of its analysis, it was proved, in particular, that general inequality \eqref{CGLinequality1} is indeed satisfied for the Robin QW. In the same way, other balanced measures can be defined and analyzed; for example, oscillatory structure and delocalization can be simultaneously defined through the following combination of the Fisher information and Shannon entropy $\frac{1}{2\pi e}I\exp\left(\frac{2}{3}S\right)$ \cite{Esquivel1}. Also, methodology presented here can be used for the analysis of more general quantum-information measures, such as R\'{e}nyi \cite{Renyi1} and Tsallis \cite{Tsallis1} entropies. We defer this investigation for future publication.

\section{Acknowledgments}
Research was supported by SEED Project No. 1702143045-P from the Research Funding Department, Vice Chancellor for Research and Graduate Studies, University of Sharjah.

\bibliographystyle{model1a-num-names}

\begin{thebibliography}{00}
\bibitem{Belloni1}M. Belloni and R. W. Robinett, \PRe {\bf 540}, 25 (2014).
\bibitem{AlHashimi1}M. H. Al-Hashimi and U.-J. Wiese, \APNY {\bf 327}, 1 (2012).
\bibitem{Gustafson1}K. Gustafson and T. Abe, {\em Math. Intell.} {\bf 20}(1), 63 (1998).
\bibitem{Olendski2}O. Olendski, \APB {\bf 527}, 278 (2015).
\bibitem{Erhart1}J. Erhart, S. Sponar, G. Sulyok, G. Badurek, M. Ozawa, and Y. Hasegawa, \NP {\bf 8}, 185 (2012).
\bibitem{Maccone1}L. Maccone and A. K. Pati, \PRL {\bf 113}, 260401 (2014).
\bibitem{Li1}T. Li, Y. Xiao, T. Ma, S.-M. Fei, N. Jing, X. Li-Jost, and Z.-X. Wang, \SR{\bf 6}, 35735 (2016).
\bibitem{Song1}Q.-C. Song and C.-F. Qiao, \PLA {\bf 380}, 2925 (2016).
\bibitem{Li2}J.-L. Li and C.-F. Qiao, \jpa{\bf 50}, 03LT01 (2017).
\bibitem{Herdegen1}A. Herdegen and P. Ziobro, \LMP{\bf 107}, 659 (2017).
\bibitem{Bialynicki2}I. Bia{\l}ynicki-Birula and {\L}. Rudnicki, in: {\em Statistical Complexity: Applications in Electronic Structure}, ed. by K. D. Sen (Springer, Dordrecht, 2011), chap. 1.
\bibitem{Bialynicki3}I. Bia{\l}ynicki-Birula and J. Mycielski, \CMP {\bf 44}, 129 (1975).
\bibitem{Beckner1}W. Beckner, \AM {\bf 102}, 159 (1975).
\bibitem{Everett1}H. Everett, III, in: {\em The Many-Worlds Interpretation of Quantum Mechanics}, ed. by B. S. DeWitt and N. Graham, Princeton Series in Physics (Princeton University Press, Princeton, 1973), chap. 1.
\bibitem{Hirschman1}I. I. Hirschman, \AJM {\bf 79}, 152 (1957).
\bibitem{Deutsch1}D. Deutsch, \PRL{\bf 50}, 631 (1983).
\bibitem{Partovi1}M. H. Partovi, \PRL{\bf 50}, 1883 (1983).
\bibitem{Wehner1}S. Wehner and A. Winter, \NJP{\bf 12}, 025009 (2010).
\bibitem{Coles1}P. J. Coles, M. Berta, M. Tomamichel, and S. Wehner, \RMP{\bf 89}, 015002 (2017).
\bibitem{Shannon1}C. E. Shannon, {\em Bell Syst. Tech. J.} {\bf 27}, 379 (1948).
\bibitem{Srinivas1}M. D. Srinivas, \PJP{\bf 24}, 673 (1985).
\bibitem{Dodonov1}V. V. Dodonov and V. I. Man'ko, {\em Tr. Fiz. Inst. Akad. Nauk SSSR} {\bf 183}, 5 (1987) [{\em Proc. Lebedev Phys. Inst., Acad. Sci. USSR} {\bf 183}, 3 (1989)].
\bibitem{Flores1}N. Flores-Gallegos, \CPL{\bf 650}, 57 (2016).
\bibitem{Sears2}S. B. Sears and S. R. Gadre, \JCP{\bf 75}, 4626 (1981).
\bibitem{Ho1}M. H\^{o}, D. F. Weaver, V. H. Smith, R. P. Sagar ,and R. O. Esquivel, \PRA{\bf 57}, 4512 (1998).
\bibitem{Esquivel2}R. O. Esquivel, N. Flores-Gallegos, C. Iuga, E. M. Carrera, J. C. Angulo, and J. Antol\'{i}n, {\em Theor. Chem. Acc.} {\bf 124}, 445 (2009).
\bibitem{Fisher1}R. A. Fisher, {\em Math. Proc. Cambridge Philos. Soc.} {\bf 22}, 700 (1925).
\bibitem{Sears1}S. B. Sears, R. B. Parr, and U. Dinur, {\em Israel J. Chem.} {\bf 19}, 165 (1980).
\bibitem{Frieden1}B. R. Frieden, {\em Science from Fisher Information} (Cambridge, Cambridge, 2004).
\bibitem{Onicescu1}O. Onicescu, \CRA {\bf 263}, 841 (1966).
\bibitem{Hyman1}A. S. Hyman, S. I. Yaniger, and J. F. Liebman, \IJQC{\bf 14}, 757 (1978).
\bibitem{Tao1}J. Tao and G. Li, \JCP{\bf 105}, 10493 (1996).
\bibitem{Stam1}A. J. Stam, {\em Inf. Control} {\bf 2}, 101 (1959).
\bibitem{Dembo1}A. Dembo, T. M. Cover, and J. A. Thomas, {\em IEEE Trans. Inf. Theory} {\bf 37}, 1501 (1991).
\bibitem{Romera1}E. Romera, P. S\'{a}nchez-Moreno, and J. S. Dehesa, \CPL{\bf 414}, 468 (2005).
\bibitem{Dehesa2}J. S. Dehesa, A. Mart\'{i}nez-Finkelshtein, and V. N. Sorokin, {\em Mol. Phys.} {\bf 104}, 613 (2006).
\bibitem{Dehesa3}J. S. Dehesa,  R. Gonz\'{a}lez-F\'{e}rez, and P. S\'{a}nchez-Moreno, \jpa{\bf 40}, 1845 (2007).
\bibitem{Ghosal1}A. Ghosal, N. Mukherjee, and A. K. Roy, \APB{\bf 528}, 796 (2016).
\bibitem{Plastino1}A. Plastino, G. Bellomo, and A. R. Plastino, {\em Adv. Math. Phys.} {\bf 2015}, 120698 (2015).
\bibitem{Saha1}A. Saha, B. Talukdar, and S. Chatterjee, \EJP{\bf 38}, 025103 (2017).
\bibitem{Esquivel1}R. O. Esquivel, S. L\'{o}pez-Rosa, M. Molina-Esp\'{i}ritu, J. C. Angulo, and J. S. Dehesa, {\em Theor. Chem. Acc.} {\bf 135}, 253 (2016).
\bibitem{Chatzisavvas1}K. C. Chatzisavvas, C. C. Moustakidis, and C. P. Panos, \JCP {\bf 123}, 174111 (2005).
\bibitem{Catalan1}R. G. Catal\'{a}n, J. Garay, and R. L\'{o}pez-Ruiz, \PRE{\bf 66}, 011102 (2002).
\bibitem{LopezRosa2}S. L\'{o}pez-Rosa, J. C. Angulo, and J. Antol\'{i}n, \PA{\bf 388}, 2081 (2009).
\bibitem{Rosso1}O. A. Rosso, M. T. Martin, and A. Plastino, \PA{\bf 320}, 497 (2003).
\bibitem{Olendski22}O. Olendski, \APB{\bf 527}, 296 (2015).
\bibitem{Olendski3}O. Olendski, \APB{\bf 528}, 865 (2016).
\bibitem{Olendski33}O. Olendski, \APB{\bf 528}, 882 (2016).
\bibitem{Olendski44}O. Olendski, \APB{\bf 2018}, {\it 530}, 1700325.
\bibitem{Felix1}S. F\'{e}lix, A. Maurel, and J.-F. Mercier, {\em Wave Motion} {\bf 54}, 1 (2015).
\bibitem{Balian1}R. Balian and C. Bloch, \APNY{\bf 60}, 401 (1970).
\bibitem{Katsenelenbaum1}B. Z. Katsenelenbaum, {\em High-Frequency Electrodynamics} (Wiley, Weinheim, Germany, 2006).
\bibitem{Silva1}H. O. Silva and C. Farina, \PRD{\bf 84}, 045003 (2011).
\bibitem{Solodukhin1}S. N. Solodukhin, \PRD{\bf 63}, 044002 (2001).
\bibitem{Saharian1}A. A. Saharian, \PRD{\bf 63}, 125007 (2001).
\bibitem{Romeo1}A. Romeo and A. A. Saharian, \jpa{\bf 35}, 1297 (2002). 
\bibitem{Elizalde1}E. Elizalde, S. D. Odintsov, and A. A. Saharian, \PRD{\bf 79}, 065023 (2009).
\bibitem{Teo1}L. P. Teo, \JHEP{\bf 2009}, 095 (2009).
\bibitem{Nazari1}B. Nazari, \APB{\bf 529}, 1700142 (2017).
\bibitem{Fink1}H. J. Fink and W. C. H. Joiner, \PRL{\bf 23}, 120 (1969).
\bibitem{Montevecchi1}E. Montevecchi and J. O. Indekeu, \EPL{\bf 51}, 661 (2000).
\bibitem{Kozhevnikov1}V. F. Kozhevnikov, M. J. Van Bael, W. Vinckx, K. Temst, C. Van Haesendonck, and J. O. Indekeu, \PRB{\bf 72}, 174510 (2005).
\bibitem{Giorgi1}T. Giorgi and R. Smit, {\em Z. Angew. Math. Phys.} {\bf 58}, 224 (2007).
\bibitem{deGennes1}P. G. de Gennes, {\em Superconductivity of Metals and Alloys} (Benjamin, New York, 1966).
\bibitem{Sapoval1}B. Sapoval, \PRL{\bf 73}, 3314 (1994).
\bibitem{Essert1}S. Essert, V. Krueckl, and K. Richter, \NJP{\bf 16}, 113058 (2014).
\bibitem{Olendski4}O. Olendski, \APNY{\bf 326}, 1479 (2011).
\bibitem{Olendski5}O. Olendski, \APNY{\bf 327}, 1365 (2012).
\bibitem{Grebenkov1}D. S. Grebenkov and B.-T. Nguyen, {\em SIAM Rev.} {\bf 55}, 601 (2013).
\bibitem{Olendski1}O. Olendski and L. Mikhailovska, \PRE{\bf 81}, 036606 (2010).
\bibitem{Lacey1}A. A. Lacey, J. R. Ockendon, and J. Sabina, {\em SIAM J. Appl. Math.} {\bf 58}, 1622 (1998).
\bibitem{Lou1}Y. Lou and M. Zhu, {\em Pacific J. Math.} {\bf 214}, 323 (2004).
\bibitem{Levitin1}M. Levitin and L. Parnovski, {\em Math. Nachr.} {\bf 281}, 272 (2008).
\bibitem{Daners1}D. Daners and J. Kennedy, {\em Differ. Integral Equ.} {\bf 23}, 659 (2010).
\bibitem{Colorado1}E. Colorado and J. Garc\'{i}a-Meli\'{a}n, {\em J. Math. Anal. Appl.} {\bf 377}, 53 (2011).
\bibitem{Pankrashkin1}K. Pankrashkin and N. Popoff, {\em J. Math. Pure. Appl.} {\bf 106}, 615 (2016).
\bibitem{Exner1}P. Exner and A. Minakov, \JMP{\bf 55}, 122101 (2014).
\bibitem{Freitas1}P. Freitas and D. Krej\v{c}i\v{r}\'{i}k, \AdM{\bf 280}, 322 (2015).
\bibitem{Slachmuylders1}A. F. Slachmuylders, B. Partoens, and F. M. Peeters, \PRB{\bf 71}, 245405 (2005).
\bibitem{Cacciapuoti1}C. Cacciapuoti and D. Finco, {\em Asymptotic Anal.} {\bf 70}, 199 (2010).
\bibitem{LopezRosa1}S. L\'{o}pez-Rosa, J. Montero, P. S\'{a}nchez-Moreno, J. Venegas, and J. S. Dehesa, \JMC{\bf 49}, 971 (2011).
\bibitem{Robinett1}R. W. Robinett, \AJP{\bf 63}, 823 (1995).
\bibitem{Majernik1}V. Majern\'{i}k and L. Richterek, \jpa{\bf 30}, L49 (1997).
\bibitem{Majernik2}V. Majern\'{i}k and L. Richterek, \EJP{\bf 18}, 79 (1997).
\bibitem{Majernik3}V. Majern\'{i}k, R. Charvot, and E. Majern\'{i}kov\'{a}, \jpa{\bf 32}, 2207 (1999).
\bibitem{Gradshteyn1}I. S. Gradshteyn and I. M. Ryzhik, {\em Table of Integrals, Series, and Products} (Academic, New York, 2014).
\bibitem{Prudnikov1}A. P. Prudnikov, Y. A. Brychkov, and O. I. Marichev, {\em Integrals and Series}, vol. 1 (Gordon and Breach, New York, 1998).
\bibitem{Brychkov1}Y. A. Brychkov, {\em Handbook Of Special Functions} (CRC Press, Boca Raton, 2008).
\bibitem{Renyi1}A. R\'{e}nyi, {\em Probability Theory} (Dover, Mineola, New York, 2007).
\bibitem{Tsallis1}C. Tsallis, \JSP {\bf 52}, 479 (1988).
\end{thebibliography}
 
\end{document}